\expandafter\ifx\csname phyzzx\endcsname\relax
 \message{It is better to use PHYZZX format than to
          \string\input\space PHYZZX}\else
 \wlog{PHYZZX macros are already loaded and are not
          \string\input\space again}%
 \endinput \fi
\catcode`\@=11 
\let\rel@x=\relax
\let\n@expand=\relax
\def\pr@tect{\let\n@expand=\noexpand}
\let\protect=\pr@tect
\let\gl@bal=\global
%
%
%
\newfam\cpfam
\newdimen\b@gheight             \b@gheight=12pt
\newcount\f@ntkey               \f@ntkey=0
\def\f@m{\afterassignment\samef@nt\f@ntkey=}
\def\samef@nt{\fam=\f@ntkey \the\textfont\f@ntkey\rel@x}
\def\setstr@t{\setbox\strutbox=\hbox{\vrule height 0.85\b@gheight
                                depth 0.35\b@gheight width\z@ }}
\input phyzzx.fonts
%
\def\rm{\n@expand\f@m0 }
\def\mit{\n@expand\f@m1 }         
\def\cal{\n@expand\f@m2 }
\def\it{\n@expand\f@m\itfam}
\def\sl{\n@expand\f@m\slfam}
\def\bf{\n@expand\f@m\bffam}
\def\tt{\n@expand\f@m\ttfam}
\def\caps{\n@expand\f@m\cpfam}    
\def\em@{\rel@x\ifnum\f@ntkey=0 \it \else
        \ifnum\f@ntkey=\bffam \it \else \rm \fi \fi }
\def\em{\n@expand\em@}
\def\fourteenpoint{\fourteenf@nts \samef@nt \b@gheight=14pt \setstr@t }
\def\twelvepoint{\twelvef@nts \samef@nt \b@gheight=12pt \setstr@t }
\def\tenpoint{\tenf@nts \samef@nt \b@gheight=10pt \setstr@t }
\normalbaselineskip = 19.2pt plus 0.2pt minus 0.1pt 
\normallineskip = 1.5pt plus 0.1pt minus 0.1pt
\normallineskiplimit = 1.5pt
\newskip\normaldisplayskip
\normaldisplayskip = 14.4pt plus 3.6pt minus 10.0pt 
\newskip\normaldispshortskip
\normaldispshortskip = 6pt plus 5pt
\newskip\normalparskip
\normalparskip = 6pt plus 2pt minus 1pt
\newskip\skipregister
\skipregister = 5pt plus 2pt minus 1.5pt
\newif\ifsingl@
\newif\ifdoubl@
\newif\iftwelv@  \twelv@true
\def\singlespace{\singl@true\doubl@false\spaces@t}
\def\doublespace{\singl@false\doubl@true\spaces@t}
\def\normalspace{\singl@false\doubl@false\spaces@t}
\def\Tenpoint{\tenpoint\twelv@false\spaces@t}
\def\Twelvepoint{\twelvepoint\twelv@true\spaces@t}
\def\spaces@t{\rel@x
      \iftwelv@ \ifsingl@\subspaces@t3:4;\else\subspaces@t1:1;\fi
       \else \ifsingl@\subspaces@t3:5;\else\subspaces@t4:5;\fi \fi
      \ifdoubl@ \multiply\baselineskip by 5
         \divide\baselineskip by 4 \fi }
\def\subspaces@t#1:#2;{
      \baselineskip = \normalbaselineskip
      \multiply\baselineskip by #1 \divide\baselineskip by #2
      \lineskip = \normallineskip
      \multiply\lineskip by #1 \divide\lineskip by #2
      \lineskiplimit = \normallineskiplimit
      \multiply\lineskiplimit by #1 \divide\lineskiplimit by #2
      \parskip = \normalparskip
      \multiply\parskip by #1 \divide\parskip by #2
      \abovedisplayskip = \normaldisplayskip
      \multiply\abovedisplayskip by #1 \divide\abovedisplayskip by #2
      \belowdisplayskip = \abovedisplayskip
      \abovedisplayshortskip = \normaldispshortskip
      \multiply\abovedisplayshortskip by #1
        \divide\abovedisplayshortskip by #2
      \belowdisplayshortskip = \abovedisplayshortskip
      \advance\belowdisplayshortskip by \belowdisplayskip
      \divide\belowdisplayshortskip by 2
      \smallskipamount = \skipregister
      \multiply\smallskipamount by #1 \divide\smallskipamount by #2
      \medskipamount = \smallskipamount \multiply\medskipamount by 2
      \bigskipamount = \smallskipamount \multiply\bigskipamount by 4 }
\def\normalbaselines{ \baselineskip=\normalbaselineskip
   \lineskip=\normallineskip \lineskiplimit=\normallineskip
   \iftwelv@\else \multiply\baselineskip by 4 \divide\baselineskip by 5
     \multiply\lineskiplimit by 4 \divide\lineskiplimit by 5
     \multiply\lineskip by 4 \divide\lineskip by 5 \fi }
\Twelvepoint  
\interlinepenalty=50
\interfootnotelinepenalty=5000
\predisplaypenalty=9000
\postdisplaypenalty=500
\hfuzz=1pt
\vfuzz=0.2pt
\newdimen\HOFFSET  \HOFFSET=0pt
\newdimen\VOFFSET  \VOFFSET=0pt
\newdimen\HSWING   \HSWING=0pt
\dimen\footins=8in
%
%
%
\newskip\pagebottomfiller
\pagebottomfiller=\z@ plus \z@ minus \z@
\def\pagecontents{
   \ifvoid\topins\else\unvbox\topins\vskip\skip\topins\fi
   \dimen@ = \dp255 \unvbox255
   \vskip\pagebottomfiller
   \ifvoid\footins\else\vskip\skip\footins\footrule\unvbox\footins\fi
   \ifr@ggedbottom \kern-\dimen@ \vfil \fi }
\def\makeheadline{\vbox to 0pt{ \skip@=\topskip
      \advance\skip@ by -12pt \advance\skip@ by -2\normalbaselineskip
      \vskip\skip@ \line{\vbox to 12pt{}\the\headline} \vss
      }\nointerlineskip}
\def\makefootline{\baselineskip = 1.5\normalbaselineskip
                 \line{\the\footline}}
\newif\iffrontpage
\newif\ifp@genum
\def\nopagenumbers{\p@genumfalse}
\def\pagenumbers{\p@genumtrue}
\pagenumbers
\newtoks\paperheadline
\newtoks\paperfootline
\newtoks\letterheadline
\newtoks\letterfootline
\newtoks\letterinfo
\newtoks\date
\paperheadline={\hfil}
\paperfootline={\hss\iffrontpage\else\ifp@genum\tenrm\folio\hss\fi\fi}
\letterheadline{\iffrontpage \hfil \else
    \rm \ifp@genum page~~\folio\fi \hfil\the\date \fi}
\letterfootline={\iffrontpage\the\letterinfo\else\hfil\fi}
\letterinfo={\hfil}
\def\monthname{\rel@x\ifcase\month 0/\or January\or February\or
   March\or April\or May\or June\or July\or August\or September\or
   October\or November\or December\else\number\month/\fi}
\def\today{\monthname~\number\day, \number\year}
\date={\today}
\headline=\paperheadline 
\footline=\paperfootline 
\countdef\pageno=1      \countdef\pagen@=0
\countdef\pagenumber=1  \pagenumber=1
\def\advancepageno{\gl@bal\advance\pagen@ by 1
   \ifnum\pagenumber<0 \gl@bal\advance\pagenumber by -1
    \else\gl@bal\advance\pagenumber by 1 \fi
    \gl@bal\frontpagefalse  \swing@ }
\def\folio{\ifnum\pagenumber<0 \romannumeral-\pagenumber
           \else \number\pagenumber \fi }
\def\swing@{\ifodd\pagenumber \gl@bal\advance\hoffset by -\HSWING
             \else \gl@bal\advance\hoffset by \HSWING \fi }
\def\footrule{\dimen@=\prevdepth\nointerlineskip
   \vbox to 0pt{\vskip -0.25\baselineskip \hrule width 0.35\hsize \vss}
   \prevdepth=\dimen@ }
\let\footnotespecial=\rel@x
\newdimen\footindent
\footindent=24pt
\def\Textindent#1{\noindent\llap{#1\enspace}\ignorespaces}
\def\Vfootnote#1{\insert\footins\bgroup
   \interlinepenalty=\interfootnotelinepenalty \floatingpenalty=20000
   \singl@true\doubl@false\Tenpoint
   \splittopskip=\ht\strutbox \boxmaxdepth=\dp\strutbox
   \leftskip=\footindent \rightskip=\z@skip
   \parindent=0.5\footindent \parfillskip=0pt plus 1fil
   \spaceskip=\z@skip \xspaceskip=\z@skip \footnotespecial
   \Textindent{#1}\footstrut\futurelet\next\fo@t}

\def\vfootnote#1{\Vfootnote{${#1}$}}
\def\footnote#1{\attach{#1}\vfootnote{#1}}

\let\footsymbol=\star
\newcount\lastf@@t           \lastf@@t=-1
\newcount\footsymbolcount    \footsymbolcount=0
\newif\ifPhysRev
\def\bumpfootsymbolcount{\rel@x
   \iffrontpage \bumpfootsymbolpos \else \advance\lastf@@t by 1
     \ifPhysRev \bumpfootsymbolneg \else \bumpfootsymbolpos \fi \fi
   \gl@bal\lastf@@t=\pagen@ }
\def\bumpfootsymbolpos{\ifnum\footsymbolcount <0
                            \gl@bal\footsymbolcount =0 \fi
    \ifnum\lastf@@t<\pagen@ \gl@bal\footsymbolcount=0
     \else \gl@bal\advance\footsymbolcount by 1 \fi }
\def\bumpfootsymbolneg{\ifnum\footsymbolcount >0
             \gl@bal\footsymbolcount =0 \fi
         \gl@bal\advance\footsymbolcount by -1 }
\def\fd@f#1 {\xdef\footsymbol{\mathchar"#1 }}
\def\generatefootsymbol{\ifcase\footsymbolcount \fd@f 13F \or \fd@f 279
        \or \fd@f 27A \or \fd@f 278 \or \fd@f 27B \else
        \ifnum\footsymbolcount <0 \fd@f{023 \number-\footsymbolcount }
         \else \fd@f 203 {\loop \ifnum\footsymbolcount >5
                \fd@f{203 \footsymbol } \advance\footsymbolcount by -1
                \repeat }\fi \fi }

\def\nonfrenchspacing{\sfcode`\.=3001 \sfcode`\!=3000 \sfcode`\?=3000
        \sfcode`\:=2000 \sfcode`\;=1500 \sfcode`\,=1251 }
\nonfrenchspacing
\newdimen\d@twidth
{\setbox0=\hbox{s.} \gl@bal\d@twidth=\wd0 \setbox0=\hbox{s}
        \gl@bal\advance\d@twidth by -\wd0 }
\def\removehglue{\loop \unskip \ifdim\lastskip >\z@ \repeat }
\def\roll@ver#1{\removehglue \nobreak \count255 =\spacefactor \dimen@=\z@
        \ifnum\count255 =3001 \dimen@=\d@twidth \fi
        \ifnum\count255 =1251 \dimen@=\d@twidth \fi
    \iftwelv@ \kern-\dimen@ \else \kern-0.83\dimen@ \fi
   #1\spacefactor=\count255 }
\def\step@ver#1{\rel@x \ifmmode #1\else \ifhmode
        \roll@ver{${}#1$}\else {\setbox0=\hbox{${}#1$}}\fi\fi }
\def\attach#1{\step@ver{\strut^{\mkern 2mu #1} }}
%
%
%
\newcount\chapternumber      \chapternumber=0
\newcount\sectionnumber      \sectionnumber=0
\newcount\equanumber         \equanumber=0
\let\chapterlabel=\rel@x
\let\sectionlabel=\rel@x
\newtoks\chapterstyle        \chapterstyle={\Number}
\newtoks\sectionstyle        \sectionstyle={\chapterlabel.\Number}
\newskip\chapterskip         \chapterskip=\bigskipamount
\newskip\sectionskip         \sectionskip=\medskipamount
\newskip\headskip            \headskip=8pt plus 3pt minus 3pt
\newdimen\chapterminspace    \chapterminspace=15pc
\newdimen\sectionminspace    \sectionminspace=10pc
\newdimen\referenceminspace  \referenceminspace=20pc
\def\chapterreset{\gl@bal\advance\chapternumber by 1
   \ifnum\equanumber<0 \else\gl@bal\equanumber=0\fi
   \sectionnumber=0 \let\sectionlabel=\rel@x
   {\pr@tect\xdef\chapterlabel{\the\chapterstyle{\the\chapternumber}}}}
\def\alphabetic#1{\count255='140 \advance\count255 by #1\char\count255}
\def\Alphabetic#1{\count255='100 \advance\count255 by #1\char\count255}
\def\Roman#1{\uppercase\expandafter{\romannumeral #1}}
\def\roman#1{\romannumeral #1}
\def\Number#1{\number #1}
\def\BLANC#1{}
\def\titleparagraphs{\interlinepenalty=9999
     \leftskip=0.03\hsize plus 0.22\hsize minus 0.03\hsize
     \rightskip=\leftskip \parfillskip=0pt
     \hyphenpenalty=9000 \exhyphenpenalty=9000
     \tolerance=9999 \pretolerance=9000
     \spaceskip=0.333em \xspaceskip=0.5em }
\def\titlestyle#1{\par\begingroup \titleparagraphs
     \iftwelv@\fourteenpoint\else\twelvepoint\fi
   \noindent #1\par\endgroup }
\def\spacecheck#1{\dimen@=\pagegoal\advance\dimen@ by -\pagetotal
   \ifdim\dimen@<#1 \ifdim\dimen@>0pt \vfil\break \fi\fi}
\def\chapter#1{\par \penalty-300 \vskip\chapterskip
   \spacecheck\chapterminspace
   \chapterreset \titlestyle{\S\chapterlabel.~#1}
   \nobreak\vskip\headskip \penalty 30000
   {\pr@tect\wlog{\string\chapter\space \chapterlabel}} }

\def\section#1{\par \ifnum\the\lastpenalty=30000\else
   \penalty-200\vskip\sectionskip \spacecheck\sectionminspace\fi
   \gl@bal\advance\sectionnumber by 1
   {\pr@tect
   \xdef\sectionlabel{\the\sectionstyle\the\sectionnumber}
   \wlog{\string\section\space \sectionlabel}}
   \noindent {\caps\enspace\sectionlabel.~~#1}\par
   \nobreak\vskip\headskip \penalty 30000 }
\def\subsection#1{\par
   \ifnum\the\lastpenalty=30000\else \penalty-100\smallskip \fi
   \noindent\undertext{#1}\enspace \vadjust{\penalty5000}}

\def\undertext#1{\vtop{\hbox{#1}\kern 1pt \hrule}}

\def\ack{\subsection{Acknowledgements:}}
\def\APPENDIX#1#2{\par\penalty-300\vskip\chapterskip
   \spacecheck\chapterminspace \chapterreset \xdef\chapterlabel{#1}
   \titlestyle{Appendix #2} \nobreak\vskip\headskip \penalty 30000
   \wlog{\string\Appendix~\chapterlabel} }
\def\Appendix#1{\APPENDIX{#1}{#1}}
\def\appendix{\APPENDIX{A}{}}
\def\unnumberedchapters{\let\makechapterlabel=\rel@x
      \let\chapterlabel=\rel@x  \sectionstyle={\BLANC}
      \let\sectionlabel=\rel@x \sequentialequations }
%
%
%
\def\eqname#1{\rel@x {\pr@tect
  \ifnum\equanumber<0 \xdef#1{{\rm(\number-\equanumber)}}%
     \gl@bal\advance\equanumber by -1
  \else \gl@bal\advance\equanumber by 1
     \ifx\chapterlabel\rel@x \def\d@t{}\else \def\d@t{.}\fi
    \xdef#1{{\rm(\chapterlabel\d@t\number\equanumber)}}\fi #1}}
\def\eq{\eqname\?}

\def\eqinsert#1{\noalign{\dimen@=\prevdepth \nointerlineskip
   \setbox0=\hbox to\displaywidth{\hfil #1}
   \vbox to 0pt{\kern 0.5\baselineskip\hbox{$\!\box0\!$}\vss}
   \prevdepth=\dimen@}}
%

%
%
\def\GENITEM#1;#2{\par \hangafter=0 \hangindent=#1
    \Textindent{$ #2 $}\ignorespaces}
\outer\def\newitem#1=#2;{\gdef#1{\GENITEM #2;}}

\newdimen\itemsize                \itemsize=30pt
\newitem\item=1\itemsize;
\newitem\sitem=1.75\itemsize;     
\newitem\ssitem=2.5\itemsize;     
\outer\def\newlist#1=#2&#3&#4;{\toks0={#2}\toks1={#3}%
   \count255=\escapechar \escapechar=-1
   \alloc@0\list\countdef\insc@unt\listcount     \listcount=0
   \edef#1{\par
      \countdef\listcount=\the\allocationnumber
      \advance\listcount by 1
      \hangafter=0 \hangindent=#4
      \Textindent{\the\toks0{\listcount}\the\toks1}}
   \expandafter\expandafter\expandafter
    \edef\c@t#1{begin}{\par
      \countdef\listcount=\the\allocationnumber \listcount=1
      \hangafter=0 \hangindent=#4
      \Textindent{\the\toks0{\listcount}\the\toks1}}
   \expandafter\expandafter\expandafter
    \edef\c@t#1{con}{\par \hangafter=0 \hangindent=#4 \noindent}
   \escapechar=\count255}
\def\c@t#1#2{\csname\string#1#2\endcsname}
\newlist\point=\Number&.&1.0\itemsize;
\newlist\subpoint=(\alphabetic&)&1.75\itemsize;
\newlist\subsubpoint=(\roman&)&2.5\itemsize;
%

%
%
%
%
\newcount\referencecount     \referencecount=0
\newcount\lastrefsbegincount \lastrefsbegincount=0
\newif\ifreferenceopen       \newwrite\referencewrite
\newdimen\refindent          \refindent=30pt
\def\normalrefmark#1{\attach{\scriptscriptstyle [ #1 ] }}
\let\PRrefmark=\attach
\def\NPrefmark#1{\step@ver{{\;[#1]}}}
\def\refmark#1{\rel@x\ifPhysRev\PRrefmark{#1}\else\normalrefmark{#1}\fi}
\def\refend@{\refmark{\number\referencecount}}
\def\refend{\refend@{}\space }
\def\refsend{\refmark{\count255=\referencecount
   \advance\count255 by-\lastrefsbegincount
   \ifcase\count255 \number\referencecount
   \or \number\lastrefsbegincount,\number\referencecount
   \else \number\lastrefsbegincount-\number\referencecount \fi}\space }
\def\REFNUM#1{\rel@x \gl@bal\advance\referencecount by 1
    \xdef#1{\the\referencecount }}
\def\Refnum#1{\REFNUM #1\refend@ } 
\def\REF#1{\REFNUM #1\R@FWRITE\ignorespaces}
\def\Ref#1{\Refnum #1\REFWRITE }
\def\ref{\Ref\?}
\def\REFS#1{\REFNUM #1\gl@bal\lastrefsbegincount=\referencecount
    \REFWRITE }

\def\r@fitem#1{\par \hangafter=0 \hangindent=\refindent \Textindent{#1}}
\def\refitem#1{\r@fitem{#1.}}
\def\NPrefitem#1{\r@fitem{[#1]}}
\def\NPrefs{\let\refmark=\NPrefmark \let\refitem=\NPrefitem}
\def\REFWRITE{\R@FWRITE\rel@x }
\def\R@FWRITE#1{\ifreferenceopen \else \gl@bal\referenceopentrue
     \immediate\openout\referencewrite=\jobname.refs
     \toks@={\begingroup \refoutspecials \catcode`\^^M=10 }%
     \immediate\write\referencewrite{\the\toks@}\fi
    \immediate\write\referencewrite{\noexpand\refitem %
                                    {\the\referencecount}}%
    \p@rse@ndwrite \referencewrite #1}
\begingroup
 \catcode`\^^M=\active \let^^M=\relax %
 \gdef\p@rse@ndwrite#1#2{\begingroup \catcode`\^^M=12 \newlinechar=`\^^M%
         \chardef\rw@write=#1\sc@nlines#2}%
 \gdef\sc@nlines#1#2{\sc@n@line \g@rbage #2^^M\endsc@n \endgroup #1}%
 \gdef\sc@n@line#1^^M{\expandafter\toks@\expandafter{\deg@rbage #1}%
         \immediate\write\rw@write{\the\toks@}%
         \futurelet\n@xt \sc@ntest }%
\endgroup
\def\sc@ntest{\ifx\n@xt\endsc@n \let\n@xt=\rel@x
       \else \let\n@xt=\sc@n@notherline \fi \n@xt }
\def\sc@n@notherline{\sc@n@line \g@rbage }
\def\deg@rbage#1{}
\let\g@rbage=\relax    \let\endsc@n=\relax
\def\refout{\par\penalty-400\vskip\chapterskip
   \spacecheck\referenceminspace
   \ifreferenceopen \Closeout\referencewrite \referenceopenfalse \fi
   \line{\fourteenrm\hfil References\hfil}\vskip\headskip
   \input \jobname.refs
   }
\def\refoutspecials{\sfcode`\.=1000 \interlinepenalty=1000
         \rightskip=\z@ plus 1em minus \z@ }
\def\Closeout#1{\toks0={\par\endgroup}\immediate\write#1{\the\toks0}%
   \immediate\closeout#1}
%
%
\newcount\figurecount     \figurecount=0
\newcount\tablecount      \tablecount=0
\newif\iffigureopen       \newwrite\figurewrite
\newif\iftableopen        \newwrite\tablewrite
\def\FIGNUM#1{\rel@x \gl@bal\advance\figurecount by 1
    \xdef#1{\the\figurecount}}
\def\FIGURE#1{\FIGNUM #1\F@GWRITE\ignorespaces }
\let\FIG=\FIGURE

\def\fig{\FIGNUM\?figure~\?\FIGWRITE }
\def\figitem#1{\r@fitem{#1)}}
\def\FIGWRITE{\F@GWRITE\rel@x }
\def\TABNUM#1{\rel@x \gl@bal\advance\tablecount by 1
    \xdef#1{\the\tablecount}}
\def\TABLE#1{\TABNUM #1\T@BWRITE\ignorespaces }

\def\tabitem#1{\r@fitem{#1:}}
\def\TABWRITE{\T@BWRITE\rel@x }
\def\F@GWRITE#1{\iffigureopen \else \gl@bal\figureopentrue
     \immediate\openout\figurewrite=\jobname.figs
     \toks@={\begingroup \catcode`\^^M=10 }%
     \immediate\write\figurewrite{\the\toks@}\fi
    \immediate\write\figurewrite{\noexpand\figitem %
                                 {\the\figurecount}}%
    \p@rse@ndwrite \figurewrite #1}
\def\T@BWRITE#1{\iftableopen \else \gl@bal\tableopentrue
     \immediate\openout\tablewrite=\jobname.tabs
     \toks@={\begingroup \catcode`\^^M=10 }%
     \immediate\write\tablewrite{\the\toks@}\fi
    \immediate\write\tablewrite{\noexpand\tabitem %
                                 {\the\tablecount}}%
    \p@rse@ndwrite \tablewrite #1}
\def\figout{\par\penalty-400
   \vskip\chapterskip\spacecheck\referenceminspace
   \iffigureopen \Closeout\figurewrite \figureopenfalse \fi
   \line{\fourteenrm\hfil FIGURE CAPTIONS\hfil}\vskip\headskip
   \input \jobname.figs
   }
\def\tabout{\par\penalty-400
   \vskip\chapterskip\spacecheck\referenceminspace
   \iftableopen \Closeout\tablewrite \tableopenfalse \fi
   \line{\fourteenrm\hfil TABLE CAPTIONS\hfil}\vskip\headskip
   \input \jobname.tabs
   }
%
%
%
\newbox\picturebox
\def\p@cht{\ht\picturebox }
\def\p@cwd{\wd\picturebox }
\def\p@cdp{\dp\picturebox }
\newdimen\xshift
\newdimen\yshift
\newdimen\captionwidth
\newskip\captionskip
\captionskip=15pt plus 5pt minus 3pt
\def\fullwidth{\captionwidth=\hsize }
\newtoks\Caption
\newif\ifcaptioned
\newif\ifselfcaptioned
\def\caption{\captionedtrue \Caption }
\newcount\linesabove
\newif\iffileexists
\newtoks\picfilename
\def\fil@#1 {\fileexiststrue \picfilename={#1}}
\def\file#1{\if=#1\let\n@xt=\fil@ \else \def\n@xt{\fil@ #1}\fi \n@xt }
\def\pl@t{\begingroup \pr@tect
    \setbox\picturebox=\hbox{}\fileexistsfalse
    \let\height=\p@cht \let\width=\p@cwd \let\depth=\p@cdp
    \xshift=\z@ \yshift=\z@ \captionwidth=\z@
    \Caption={}\captionedfalse
    \linesabove =0 \picturedefault }
\def\plot{\pl@t \selfcaptionedfalse }
\def\Picture#1{\gl@bal\advance\figurecount by 1
    \xdef#1{\the\figurecount}\pl@t \selfcaptionedtrue }

\def\s@vepicture{\iffileexists \parsefilename \redopicturebox \fi
   \ifdim\captionwidth>\z@ \else \captionwidth=\p@cwd \fi
   \xdef\lastpicture{\iffileexists
        \setbox0=\hbox{\raise\the\yshift \vbox{%
              \moveright\the\xshift\hbox{\picturedefinition}}}%
        \else \setbox0=\hbox{}\fi
         \ht0=\the\p@cht \wd0=\the\p@cwd \dp0=\the\p@cdp
         \vbox{\hsize=\the\captionwidth \line{\hss\box0 \hss }%
              \ifcaptioned \vskip\the\captionskip \noexpand\Tenpoint
                \ifselfcaptioned Figure~\the\figurecount.\enspace \fi
                \the\Caption \fi }}%
    \endgroup }
\let\endpicture=\s@vepicture
\def\savepicture#1{\s@vepicture \global\let#1=\lastpicture }
\def\displaypicture{\fullwidth \s@vepicture $$\lastpicture $${}}
\def\toppicture{\fullwidth \s@vepicture \topinsert
    \lastpicture \medskip \endinsert }
\def\midpicture{\fullwidth \s@vepicture \midinsert
    \lastpicture \endinsert }
%
%
\def\leftpicture{\pres@tpicture
    \dimen@i=\hsize \advance\dimen@i by -\dimen@ii
    \setbox\picturebox=\hbox to \hsize {\box0 \hss }%
    \wr@paround }
\def\rightpicture{\pres@tpicture
    \dimen@i=\z@
    \setbox\picturebox=\hbox to \hsize {\hss \box0 }%
    \wr@paround }
\def\pres@tpicture{\gl@bal\linesabove=\linesabove
    \s@vepicture \setbox\picturebox=\vbox{
         \kern \linesabove\baselineskip \kern 0.3\baselineskip
         \lastpicture \kern 0.3\baselineskip }%
    \dimen@=\p@cht \dimen@i=\dimen@
    \advance\dimen@i by \pagetotal
    \par \ifdim\dimen@i>\pagegoal \vfil\break \fi
    \dimen@ii=\hsize
    \advance\dimen@ii by -\parindent \advance\dimen@ii by -\p@cwd
    \setbox0=\vbox to\z@{\kern-\baselineskip \unvbox\picturebox \vss }}
\def\wr@paround{\Caption={}\count255=1
    \loop \ifnum \linesabove >0
         \advance\linesabove by -1 \advance\count255 by 1
         \advance\dimen@ by -\baselineskip
         \expandafter\Caption \expandafter{\the\Caption \z@ \hsize }%
      \repeat
    \loop \ifdim \dimen@ >\z@
         \advance\count255 by 1 \advance\dimen@ by -\baselineskip
         \expandafter\Caption \expandafter{%
             \the\Caption \dimen@i \dimen@ii }%
      \repeat
    \edef\n@xt{\parshape=\the\count255 \the\Caption \z@ \hsize }%
    \par\noindent \n@xt \strut \vadjust{\box\picturebox }}
\let\picturedefault=\relax
\let\parsefilename=\relax
\def\redopicturebox{\let\picturedefinition=\rel@x
   \errhelp=\disabledpictures
   \errmessage{This version of TeX cannot handle pictures.  Sorry.}}
\newhelp\disabledpictures
     {You will get a blank box in place of your picture.}
%
%
%
%
%
%
%
%
%
%
\def\FRONTPAGE{\ifvoid255\else\vfill\penalty-20000\fi
   \gl@bal\pagenumber=1     \gl@bal\chapternumber=0
   \gl@bal\equanumber=0     \gl@bal\sectionnumber=0
   \gl@bal\referencecount=0 \gl@bal\figurecount=0
   \gl@bal\tablecount=0     \gl@bal\frontpagetrue
   \gl@bal\lastf@@t=0       \gl@bal\footsymbolcount=0}

\def\papers{\papersize\headline=\paperheadline\footline=\paperfootline}
\def\papersize{
   \advance\hoffset by\HOFFSET \advance\voffset by\VOFFSET
   \pagebottomfiller=0pc
   \skip\footins=\bigskipamount \normalspace }
\papers  
%
%
\newskip\lettertopskip       \lettertopskip=20pt plus 50pt
\newskip\letterbottomskip    \letterbottomskip=\z@ plus 100pt
\newskip\signatureskip       \signatureskip=40pt plus 3pt
\def\lettersize{\hsize=6.5in \vsize=8.5in \hoffset=0in \voffset=0.5in
   \advance\hoffset by\HOFFSET \advance\voffset by\VOFFSET
   \pagebottomfiller=\letterbottomskip
   \skip\footins=\smallskipamount \multiply\skip\footins by 3
   \singlespace }
\def\MEMO{\lettersize \headline=\letterheadline \footline={\hfil }%
   \let\rule=\memorule \FRONTPAGE \memohead }

\def\memodate{\afterassignment\MEMO \date }
\def\memit@m#1{\smallskip \hangafter=0 \hangindent=1in
    \Textindent{\caps #1}}
\def\subject{\memit@m{Subject:}}
\def\topic{\memit@m{Topic:}}
\def\from{\memit@m{From:}}
\def\memorule{\medskip\hrule height 1pt\bigskip}  
\def\memohead{\centerline{\fourteenrm MEMORANDUM}}
\newwrite\labelswrite
\newtoks\rw@toks
\def\letters{\lettersize
   \headline=\letterheadline \footline=\letterfootline
   \immediate\openout\labelswrite=\jobname.lab}

\let\letterhead=\rel@x
\def\addressee#1{\medskip\line{\hskip 0.75\hsize plus\z@ minus 0.25\hsize
                               \the\date \hfil }%
   \vskip \lettertopskip
   \ialign to\hsize{\strut ##\hfil\tabskip 0pt plus \hsize \crcr #1\crcr}
   \writelabel{#1}\medskip \noindent\hskip -\spaceskip \ignorespaces }
\def\rwl@begin#1\cr{\rw@toks={#1\crcr}\rel@x
   \immediate\write\labelswrite{\the\rw@toks}\futurelet\n@xt\rwl@next}
\def\rwl@next{\ifx\n@xt\rwl@end \let\n@xt=\rel@x
      \else \let\n@xt=\rwl@begin \fi \n@xt}
\let\rwl@end=\rel@x
\def\writelabel#1{\immediate\write\labelswrite{\noexpand\labelbegin}
     \rwl@begin #1\cr\rwl@end
     \immediate\write\labelswrite{\noexpand\labelend}}
\newtoks\FromAddress         \FromAddress={}
\newtoks\sendername          \sendername={}
\newbox\FromLabelBox
\newdimen\labelwidth          \labelwidth=6in
\def\makelabels{\afterassignment\Makelabels \sendersname=}
\def\Makelabels{\FRONTPAGE \letterinfo={\hfil } \MakeFromBox
     \immediate\closeout\labelswrite  \input \jobname.lab\vfil\eject}
\let\labelend=\rel@x
\def\labelbegin#1\labelend{\setbox0=\vbox{\ialign{##\hfil\cr #1\crcr}}
     \MakeALabel }
\def\MakeFromBox{\gl@bal\setbox\FromLabelBox=\vbox{\Tenpoint
     \ialign{##\hfil\cr \the\sendername \the\FromAddress \crcr }}}
\def\MakeALabel{\vskip 1pt \hbox{\vrule \vbox{
        \hsize=\labelwidth \hrule\bigskip
        \leftline{\hskip 1\parindent \copy\FromLabelBox}\bigskip
        \centerline{\hfil \box0 } \bigskip \hrule
        }\vrule } \vskip 1pt plus 1fil }
\def\signed#1{\par \nobreak \bigskip \dt@pfalse \begingroup
  \everycr={\noalign{\nobreak
            \ifdt@p\vskip\signatureskip\gl@bal\dt@pfalse\fi }}%
  \tabskip=0.5\hsize plus \z@ minus 0.5\hsize
  \halign to\hsize {\strut ##\hfil\tabskip=\z@ plus 1fil minus \z@\crcr
          \noalign{\gl@bal\dt@ptrue}#1\crcr }%
  \endgroup \bigskip }
\newbox\letterb@x
\def\lettertext{\par \vskip\parskip \unvcopy\letterb@x \par }
\def\multiletter{\setbox\letterb@x=\vbox\bgroup
      \everypar{\vrule height 1\baselineskip depth 0pt width 0pt }
      \singlespace \topskip=\baselineskip }
\def\letterend{\par\egroup}
%
%
%
\newskip\frontpageskip
\newtoks\Pubnum   
\newtoks\Pubtype  \let\pubtype=\Pubtype
\newif\ifp@bblock  \p@bblocktrue
\def\PH@SR@V{\doubl@true \baselineskip=24.1pt plus 0.2pt minus 0.1pt
             \parskip= 3pt plus 2pt minus 1pt }
\def\PHYSREV{\papers\PhysRevtrue\PH@SR@V}

\def\titlepage{\FRONTPAGE\papers\ifPhysRev\PH@SR@V\fi
   \ifp@bblock\p@bblock \else\hrule height\z@ \rel@x \fi }
\def\nopubblock{\p@bblockfalse}
\def\endpage{\vfil\break}
\frontpageskip=12pt plus .5fil minus 2pt
\Pubtype={}
\Pubnum={}
\def\p@bblock{\begingroup \tabskip=\hsize minus \hsize
   \baselineskip=1.5\ht\strutbox \topspace-2\baselineskip
   \halign to\hsize{\strut ##\hfil\tabskip=0pt\crcr
       \the\Pubnum\crcr\the\date\crcr\the\pubtype\crcr}\endgroup}
\def\title#1{\vskip\frontpageskip \titlestyle{#1} \vskip\headskip }
\def\author#1{\vskip\frontpageskip\titlestyle{\twelvecp #1}\nobreak}

\def\address#1{\par\kern 5pt\titlestyle{\twelvepoint\it #1}}
\def\andaddress{\par\kern 5pt \centerline{\sl and} \address}

\def\abstract{\par\dimen@=\prevdepth \hrule height\z@ \prevdepth=\dimen@
   \vskip\frontpageskip\centerline{\fourteenrm Abstract}\vskip\headskip }

%
%
%

\def\\{\rel@x \ifmmode \backslash \else {\tt\char`\\}\fi }
\def\sequentialequations{\rel@x \if\equanumber<0 \else
  \gl@bal\equanumber=-\equanumber \gl@bal\advance\equanumber by -1 \fi }
\def\journal#1&#2(#3){\begingroup \let\journal=\dummyj@urnal
    \unskip, \sl #1\unskip~\bf\ignorespaces #2\rm
    (\afterassignment\j@ur \count255=#3), \endgroup\ignorespaces }
\def\j@ur{\ifnum\count255<100 \advance\count255 by 1900 \fi
          \number\count255 }
\def\dummyj@urnal{%
    \toks@={Reference foul up: nested \journal macros}%
    \errhelp={Your forgot & or ( ) after the last \journal}%
    \errmessage{\the\toks@ }}

\def\topspace{\hrule height 0pt depth 0pt \vskip}

\def\Buildrel#1\under#2{\mathrel{\mathop{#2}\limits_{#1}}}
\def\becomes#1{\mathchoice{\becomes@\scriptstyle{#1}}
   {\becomes@\scriptstyle{#1}} {\becomes@\scriptscriptstyle{#1}}
   {\becomes@\scriptscriptstyle{#1}}}
\def\becomes@#1#2{\mathrel{\setbox0=\hbox{$\m@th #1{\,#2\,}$}%
        \mathop{\hbox to \wd0 {\rightarrowfill}}\limits_{#2}}}
\def\Tr{\mathop{\rm Tr}\nolimits}

\let\int=\intop         \let\oint=\ointop
\def\lsim{\mathrel{\mathpalette\@versim<}}
\def\gsim{\mathrel{\mathpalette\@versim>}}
\def\@versim#1#2{\vcenter{\offinterlineskip
        \ialign{$\m@th#1\hfil##\hfil$\crcr#2\crcr\sim\crcr } }}
\def\big#1{{\hbox{$\left#1\vbox to 0.85\b@gheight{}\right.\n@space$}}}
\def\Big#1{{\hbox{$\left#1\vbox to 1.15\b@gheight{}\right.\n@space$}}}
\def\bigg#1{{\hbox{$\left#1\vbox to 1.45\b@gheight{}\right.\n@space$}}}
\def\Bigg#1{{\hbox{$\left#1\vbox to 1.75\b@gheight{}\right.\n@space$}}}
\def\){\mskip 2mu\nobreak }
%
%
%
\let\sec@nt=\sec
\def\sec{\rel@x\ifmmode\let\n@xt=\sec@nt\else\let\n@xt\section\fi\n@xt}
\def\obsolete#1{\message{Macro \string #1 is obsolete.}}
\def\firstsec#1{\obsolete\firstsec \section{#1}}
\def\firstsubsec#1{\obsolete\firstsubsec \subsection{#1}}
\def\thispage#1{\obsolete\thispage \gl@bal\pagenumber=#1\frontpagefalse}
\def\thischapter#1{\obsolete\thischapter \gl@bal\chapternumber=#1}
\def\splitout{\obsolete\splitout\rel@x}
\def\prop{\obsolete\prop \propto }
\def\nextequation#1{\obsolete\nextequation \gl@bal\equanumber=#1
   \ifnum\the\equanumber>0 \gl@bal\advance\equanumber by 1 \fi}
\def\BOXITEM{\afterassigment\B@XITEM\setbox0=}
\def\B@XITEM{\par\hangindent\wd0 \noindent\box0 }
%
%
%
\def\phyzzx{PHY\setbox0=\hbox{Z}\copy0 \kern-0.5\wd0 \box0 X}
        
\everyjob{\xdef\today{\monthname~\number\day, \number\year}
        \input myphyx.tex }
\message{ by V.K.}
%
\catcode`\@=12 
%

\sequentialequations

\Pubnum{EPHOU-95-003(Revised2)}
\date{ }
\titlepage
\title{Integer Quantum Hall Effect with Realistic Boundary Condition : 
Exact Quantization and Breakdown.}
\author{\rm K. Ishikawa$^a$, N. Maeda$^a$, and K. Tadaki$^b$}

\address{$^a$Department of Physics, Hokkaido University, Sapporo 060, 
Japan}
\address{$^b$Department of Mathematics, Hokkaido University, Sapporo 060, 
Japan}

\abstract

A theory of integer quantum Hall effect(QHE) in realistic systems 
based on von Neumann lattice is presented. 
We show that the momentum representation is quite useful and that 
the quantum Hall regime(QHR), which is defined by the propagator 
in the momentum representation, is realized. 
In QHR, 
the Hall conductance is given by a topological invariant of the 
momentum space and is quantized exactly. 
The edge states do not modify the value and topological property 
of $\sigma_{xy}$ in QHR. 
We next compute distribution of current based on effective action 
and find a finite amount of current in the bulk and the edge, generally. 
Due to the Hall electric field in the bulk, 
breakdown of the QHE occurs. 
The critical electric field of the breakdown 
is proportional to $B^{3/2}$ and the 
proportional constant has no dependence on Landau levels 
in our theory, in agreement with the recent experiments.

\endpage

\doublespace

\hsize=469pt

\REF\a{N. Imai, K. Ishikawa, T. Matsuyama, and I. Tanaka, 
Phys. Rev. {\bf B42}, 10610 (1990); 
See also: K. Ishikawa and T. Matsuyama, Z. Phys. {\bf C33}, 41 (1986).}
\REF\b{K. Ishikawa, 
Prog. Theor. Phys. Suppl. {\bf 107}, 167 (1992).}
\REF\c{K. Ishikawa, N. Maeda, and K. Tadaki, 
Phys. Rev. {\bf B51}, 5048 (1995).}
\REF\ca{B. I. Halperin, Phys. Rev. Lett. {\bf 25}, 2185 (1982); 
R. B. Laughlin, Phys. Rev. {\bf B23}, 56 (1982).}
\REF\d{Q. Niu and D. J. Thouless, 
Phys. Rev. {\bf B35}, 2188 (1987); See also, 
Q. Niu, D. J. Thouless, and Y. S. Wu, 
Phys. Rev. {\bf B31}, 3372 (1985).}
\REF\e{K. v Klitzing, G. Dorda, and M. Pepper, 
Phys. Rev. Lett. {\bf 45}, 494 (1980);
S. Kawaji and J. Wakabayashi, in {\it Physics in High Magnetic 
Fields}, edited by S. Chikazumi and N. Miura(Springer-Verlag, Berlin 
1981).}
\REF\f{M. B\"uttiker, 
Phys. Rev. {\bf B38}, 9375 (1988);
R. Landauer, 
IBM J. Res. Dev. {\bf 1}, 223 (1957);
See also, 
H. Bavanger and D. Stone, Phys. Rev. {\bf B40}, 8169 (1989); 
M. B\"uttiker, Phys. Rev. {\bf B46}, 12485 (1992).}
\REF\g{S. Kawaji, K. Hirakawa, and M. Nagata, 
Physica {\bf B184}, 17 (1993);
G. Ebert et al, 
J. Phys. {\bf C16}, 5441 (1983); 
M. E. Cage et al, Phys. Rev. Lett. {\bf 51}, 1374 (1983); 
See also, N. Q. Balaban, U. Meirav. and H. Shtrikman, Phys. Rev. 
{\bf B52}, R5503 (1995).}
\REF\h{Some unusual properties of von Neumann lattice have been pointed 
out recently by F. Low, Complete Sets of Wave-Packets, in {\it 
Passion for Physics : Essays in Honor of Gef. Chew}(World-Sientific, 
Singapole, 1985), p.17. 
Ours is irrelevant to them, because two sets of variables are involved 
and dual basis is used. 
Eq.(1.11) shows an example of usefulness of our representation. 
See also Ref.3, and, E. Brown, Phys. Rev. {\bf 133}, 1038 (1964); 
I. Dana and I. Zak, Phys. Rev. {\bf B28}, 811 (1983); 
D. J. Thouless, J. Phys. {\bf C17}, L325 (1984).}
\REF\i{R. Kubo, S. J. Miyake, and N. Hashitsume, in {\it Solid State 
Physics}, edited by F. Seitz and D. Turnbull(Academic, New York, 1965), 
Vol. 17, p.269; 
T. Ando, Y. Matsumoto, and Y. Uemura, 
J. Phys. Soc. Jpn. {\bf 39}, 272 (1975).}
\REF\j{See Appendix of Ref.1.}
\REF\k{S. Coleman and B. Hill, 
Phys. Lett. {\bf 159B}, 184 (1985).}
\REF\l{In the presence of vector potential, a gauge transformation 
has chiral anomaly term. 
This will be addressed in Section 6 when gauge invariant effective 
potential is obtained; 
See also, X. G. Wen, Phys. Rev. Lett. {\bf 64}, 2206 (1990); 
M. Stone, Ann. Phys. (N.Y.) {\bf 207}, 38 (1991).}
\REF\m{It would be interesting to see that the exact quantization 
of $\sigma_{xy}$ as $e^2/h$ leads to the linear relation between 
the voltage and the current.}
\REF\n{Three-dimensional Chern-Simons action in finite region is not 
gauge invariant, but becomes invariant if it is combined with a 
two-dimensional non-local term which comes from chiral edge states. 
Two-dimensional chiral anomaly induces this term. 
The electric current has no contribution from this 
two-dimensional term. 
For a detail, see, N. Maeda, ``Chiral Anomaly and Effective Field 
Theory for the Quantum Hall Liquid with Edges", EPHOU-95-004
(to be published in Physics Letters B).}
\REF\o{K. Ishikawa, 
Phys. Rev. Lett. {\bf 53}, 1615 (1984).}
\REF\p{It is interesting to compare finite temperature effects of 
4D anomaly and 3D anomaly. 
H. Itoyama and A. H. Mueller, 
Nucl. Phys. {\bf B218}, 349 (1983); 
K. Ishikawa and T. Matsuyama, 
Nucl. Phys. {\bf B280}, 523 (1987).}
\REF\pq{A. H. MacDonald, T. M. Rice, and W. F. Brinkman, 
Phys. Rev. {\bf B 28}, 3648 (1983); 
D. J. Thouless, J. Phys. {\bf C 18}, 6211 (1985); 
C. W. J. Beenakker and H. van Houten, in {\it Solid State Physics}, 
edited by H. Ehrenreich and D. Turnbull(Academic, New York, 1992), 
Vol. 44, p.1.} 
\REF\pqr{A. Usher, R. J. Nicholas, J. J. Harris, and 
C. T. Foxton, Phys. Rev. {\bf B41}, 1129 (1990).}
\REF\rt{L. Eaves and F. W. Sheard, Semicond. Sci. Technol. 
{\bf 1}, 346 (1986).}
\REF\rs{S. A. Trugman, Phys. Rev. {\bf B27}, 7539 (1985); 
V. N. Nicopoulos and S. A. Trugman, Phys. Rev. Lett. 
{\bf 65}, 779 (1990); 
See also, S. A. Trugman and F. R. Waugh, Surf. Sci. {\bf 196}, 
171 (1988).}

\chapter{Introduction}

We study the quantum Hall effect in realistic systems in the 
present paper. 
Especially finite size effects and finite current effects are 
investigated in details. 

Two-dimensional electrons in a strong perpendicular magnetic field have
discrete energies with a magnetic field dependent 
finite degeneracy per area at each energy. 
When degenerate electrons are distributed uniformly, distance between two
electrons has a minimum value. 
Conjugate momentum, hence, is defined on a torus. 
Thus the momentum space is compact. 
The energy space seems not to be compact but physical requirement to the 
propagator at $p_0=\pm\infty$ or $\pm i\infty$ leads the energy space 
defined by the propagator compact. 
A representation of having these properties in a manifest manner was 
constructed based on von Neumann lattice of guiding center coordinates, 
and exact low energy theorems were given based on this representation
$^{\a,\b}$. 
Namely, it was shown that Hall conductance 
$\sigma_{xy}$ is a topological invariant of a mapping from 
the momentum space to a space defined by the propagator, and the 
quantization of the $\sigma_{xy}$ as $(e^2/h)N$ at the plateau, 
thus, is exact 
in systems of interactions and disorders. 
One-particle properties in systems of short range impurities, boundary 
potential, and periodic array of potentials have been obtained
$^{\c}$ 
also based on the same representation. 
Localization is shown easily in our method. 

From our previous investigations, the 
$\sigma_{xy}$ becomes a topological invariant of the momentum space 
and is quantized as $(e^2/h)N$ exactly at the plateau if all the 
one-particle states around Fermi energy are either localized or have 
finite energy gap. 
Hereafter we call this energy region as quantum Hall regime(QHR). 
In finite system with boundary, there are edge states$^{\ca}$ 
which are extended 
along the edge and have continuous energy across the Fermi energy, 
generally. 
They may give a finite correction to the quantized value. 
We will show in the present work that a resummation formula of the 
$\sigma_{xy}$ with 
the momentum representation is valid in finite systems as well 
and QHR is defined based on the propagator in the momentum representation. 
The outside regions of the continuous band 
of propagator in the resummation formula agrees with QHR. 
Perturbative series converges well in QHR. 
Consequently, a finite size correctoin disappears in QHR. 
Niu and Thouless$^{\d}$ 
claimed before that the finite size 
correction is of exponential type. 
Our results show that even an exponential correction does not exist 
in QHR. 

The quantized Hall conductance$^{\e}$ 
is used as a standard of resistance and to 
determine fine structure constant, $\alpha$. 
Its precise value is needed in testing quantum electorodynamics. 
The theoretical foundation of the quantum Hall effect(QHE) 
is, however, still insufficient. 
Especially there is a controversy in current distribution in a finite 
system and in finite size corrections of the QHE. 
Breakdown of the QHE is also observed recently if the current becomes 
large. 

Experiments are done with semi-conductors of Hall bar geometry. 
There are edges in such systems. 
Concerning edge states, a controversy is the following. 
In one method for a proof of QHE, 
B\"uttiker-Landauer$^{\f}$ formula is used. 
It is assumed that the one-dimensional edge states are the 
only current carrying
states around the Fermi energy and are connected with the leads. 
The quantization of the Hall conductance 
due to the one-dimensional edge states was derived. 
A correction may be caused by backscattering of edge states. 
In other approaches, the bulk states carry the currents and their roles 
are impotant. 
The edge states might give a correction to the quantized value, 
because they are extended in one direction and have continuous energies. 
The role of edge states and bulk states are thus reversed. 
Two approaches are thus very different. 
It is a purpose of the present work to resolve these controversies and 
to give a foundation of the quantum Hall effect in finite systems. 

We study a finite size effect and a finite current effect 
based on von Neumann lattice representation in the 
present paper. 
We show that the momentum representation 
and resummations of the diagrams based on the momentum representations 
are valid in finite systems and the QHR is defined as the outside region 
of the continuous energy band using the resummation formula of 
the $\sigma_{xy}$. 
It is shown that 
the quantization is exact in QHR of 
finite systems despite the fact that the edge states have continuous 
energies and cross Fermi energy. 
There are current flow both in the bulk and near the edges generally and 
due to the Hall electric field at the bulk, the QHR becomes 
narrower as the current becomes larger and eventually disappears at a 
critical value. 
QHE is broken, then. 
We estimate a critical Hall electric field 
and find an agreement with the experiments. 

The paper is organized in the following manner. 
We review our representation of two-dimensional electrons in the magnetic 
field based on magnetic von Neumann lattice in the rest of Section 1. 
In Section 2, one-particle properties are studied. 
Exact low energy 
theorem concerning the slope of current correlation function in 
infinite system are given in Section 3. 
Topological invariant expression of Hall conductance is given 
and the corrections due to interactions and disorders are shown to 
disappear in QHR. 
Finite system with cylinder geometry is 
discussed in Section 4 and finite 
systems with Hall bar geometry, which is the geometry of realistic 
experiments, are discussed in Section 5. 
In Section 6, the current distribution is computed with a use of 
effective potential. 
Breakdown of QHE is also discussed. 
A critical value of Hall electric field 
from our theory 
is proportional to $B^{3/2}$ and 
the proportional constant is independent of Landau levels. 
These properties are in agreement with the recent experiments by 
Kawaji et al$^\g$. 
Summary is given in Section 7. 

We review our representation$^{\a,\b,\c}$ 
of two-dimensional electrons in a strong 
perpendicular magnetic field here. 
It is based on von Neumann magnetic lattice$^{\h}$ 
and has excellent properties 
for a purpose of studying one-particle properties 
and of giving the exact low energy theorem of quantum field theory 
in quantum Hall system. 
Localized states are studied with the coordinate representation 
and extended states are studied with the momentum representation. 

A one-body Hamiltonian of a planar charged particle with a strong
perpendicular magnetic field, 

$$
H={(\vec p+ e\vec A)^2\over 2m},\ \partial_1 A_2-\partial_2 A_1=B,
\eqno\eq
$$
is expressed as 
$$
H={e^2 B^2\over 2m} (\xi^2+\eta^2),
\eqno\eq
$$
with two-dimensional relative coordinates in the magnetic field
$^{\i}$. 
They are defined by 
$$
\eqalign{
&\xi={1\over eB}(p_y+eA_y),\cr
&\eta=-{1\over eB}(p_x+eA_x),\cr
&[\xi,\eta]=-{i\hbar\over eB},
}
\eqno\eq
$$ 
and commute with another set of coordinates, center coordinates, $(X,Y)$ 
defined by
$$
\eqalign{
&X=x-\xi,\cr
&Y=y-\eta,\cr
&[X,Y]={i\hbar\over eB},\cr
[\xi,X]&=[\xi,Y]=[\eta,X]=[\eta,Y]=0.
}
\eqno\eq
$$
It is convenient to use variables $(X,Y)$ and $(\xi,\eta)$, instead of 
$(x,y)$ for studying the electrons in the magnetic field. 
A set of well localized functions, 
$$
\eqalign{
&f_l(\xi,\eta)\otimes\vert R_{m,n}\rangle,\cr
&f_l(\xi,\eta)\otimes\langle \tilde R_{m,n}\vert,
}
\eqno\eq
$$
defined by 
$$
\eqalign{
H_0f_l(\xi,\eta)&=E_l f_l(\xi,\eta),\ E_l=
{\hbar eB\over m}(l+{1\over2}),\cr
\vert R_{m,n}\rangle&=(-1)^{mn+m+n}e^{A^\dagger\sqrt{\pi}(m+in)
-A\sqrt{\pi}(m-in)}\vert0\rangle,\cr
\langle R_{m_1,n_1}\vert R_{m_2,n_2}\rangle&=
e^{i\pi[(m_1-m_2+1)(n_1-n_2+1)-1]}
e^{ -(\pi/2)[(m_1-m_2)^2+(n_1-n_2)^2]},\cr
\sum_{m_1,n_1}\langle R_{m_1,n_1}\vert R_{m_2,n_2}\rangle&=
\sum_{m_2,n_2}\langle R_{m_1,n_1}\vert R_{m_2,n_2}\rangle=0,\cr
R_{m,n}=a(m,n)&,\ m,n:{\rm integer},\ a=\sqrt{2\pi\hbar\over eB},\cr
A\vert0\rangle=0,&\ A=\sqrt{eB\over 2\hbar}(X+iY),\cr
A\vert R_{m,n}\rangle&=\sqrt{\pi}(m+in)\vert R_{m,n}\rangle,
}
\eqno\eq
$$
is a complete set and is used as base functions. 
Dual basis $\langle \tilde R_{m,n}\vert$ is defined, with a Green's 
function $G(m_1,n_1;m_2,n_2)$, by
$$
\eqalign{
&\langle \tilde R_{m_1,n_1}\vert=\sum_{m_2,n_2}G(m_1,n_1;m_2,n_2)
\langle R_{m_2,n_2}\vert,\cr
&\sum_{m',n'}G(m_1,n_1;m',n')\langle R_{m',n'}\vert R_{m_2,n_2}\rangle
=\delta_{m_1,m_2}\delta_{n_1,n_2}-1/N,\cr
&\sum_{m,n}=N.
}
\eqno\eq
$$
$G(m_1,n_1;m_2,n_2)$ is well localized in small $\vert m_1-m_2\vert$ 
and $\vert n_1-n_2\vert$ region as is shown in Ref.(3). 
We expand the electron field as, 
$$
\eqalign{
\Psi&=\sum_{l,m,n}a_l(m,n)f_l(\xi,\eta)\otimes\vert R_{m,n}\rangle,\cr
\Psi^\dagger
&=\sum_{l,m,n}b_l(m,n)f_l(\xi,\eta)\otimes\langle\tilde R_{m,n}\vert,\cr
}
\eqno\eq
$$
and regard coefficients $a_l(m,n)$ and $b_l(m,n)$ as quantized operators. 
The free kinetic term, potential term, and the electromagnetic current are 
expressed with these new variables as
$$
\eqalign{
&\int d\vec x\Psi^\dagger(x){(\vec p+e\vec A)^2\over 2m}\Psi(x)=
\sum_{l,m,n}E_lb(m,n)a(m,n),\cr
&\int d\vec x\Psi^\dagger(x)\Psi(x)V(x)=\sum b_{l_1}(m_1,n_1)a_{l_2}(m_2,n_2)
V_{l_1,l_2}(m_1,n_1;m_2,n_2),\cr
&J_\mu(x)=\sum b_{l_1}(m_1,n_1)a_{l_2}(m_2,n_2)
\Gamma_\mu^{l_1,l_2}(m_1,n_1;m_2,n_2;x),\cr
&V_{l_1,l_2}(m_1,n_1;m_2,n_2)=\int{d^2k\over(2\pi)^2}
\langle\tilde R_{m_1,n_1}\vert e^{i\vec k\vec X}\vert R_{m_2,n_2}\rangle
(f_{l_1}e^{i\vec k\vec\xi}f_{l_2})V(\vec k),\cr
&V(\vec x)=\int {d^2k\over(2\pi)^2}e^{i\vec k\vec x}V(\vec k),\cr
&\Gamma_\mu^{l_1,l_2}(m_1,n_1;m_2,n_2;x)=\int {d^2k\over(2\pi)^2}
\langle\tilde R_{m_1,n_1}\vert e^{i\vec k\vec X}\vert R_{m_2,n_2}\rangle
(f_{l_1}\xi_\mu e^{i(k_x \xi+k_y\eta)}f_{l_2})e^{-i\vec k\vec x},\cr
&\xi_\mu=(1,-(eB/m)\eta,(eB/m)\xi),\cr
}
\eqno\eq
$$
Advantages of the present representation are summarized in the following:

(1) The kinetic term is diagonal in the Landau level index and 
two-dimensional lattice coordinates and only the potential term has off 
diagonal terms. 
Hence the localization by short range random impurities in energy regions 
$\vert E-E_l\vert>\delta$ with a small impurity dependent parameter 
$\delta$ are shown easily in the present method. 

(2) Functions used as basis and dual basis 
are well localized around center coordinate 
$R_{m,n}$. 
Hence the multi-pole expansion of current operator itself and of commutation 
relations between the current and field operators are applicable. 
Due to these properties it is a straightforward matter to derive 
Ward-Takahashi identity and related identities as well as the exact low 
energy theorem. 
Coordinate dependence of a potential $V(\vec x)$ is preserved in 
$V_{l_1,l_2}(m_1,m_2;n_1,n_2)^\h$ with a smeared form. 
For instance, a short range potential, 
$$
V(\vec x)=g\delta (x), 
\eqno\eq
$$
is transformed into 
$$
V_{0,0}(m_1,m_2;n_1,n_2)={g\over a^2}e^{-{\pi\over 2}(
m_1^2+n_1^2+m_2^2+n_2^2)}[1+O(e^{-{\pi\over 2}})],
\eqno\eq
$$
which is a short range potential with an extention of few magnetic 
distance. 
Other components $V_{l_1,l_2}(m_1,m_2;n_1,n_2)$ have the same 
properties.

(3) Apart from disorder potentials, translational invariance is manifest 
and Fourier transformation of the operators can be defined, because 
magnetic translations commute each other and constitute abelian group 
in our lattice system. 
Extended states can be studied with the momentum representation. 

(4) The coordinates are defined on lattice sites. 
Since the momentum defined by Fourier transformation from lattice sites 
is defined on a torus, which is compact, the $\sigma_{xy}$ becomes 
a topological invariant in momentum space manifestly. 
Neither artificial boundary condition in configuration space nor 
periodic potential are needed. 

\chapter{One-particle states in finite systems with disorders and 
interactions}

We study one-particle properties of systems with one short range impurity, 
dilute short range impurities, boundary potential, and interactions 
between electrons in this section. 
Localized states are studied with the coordinate representation 
and the extended states are studied with the momentum representation. 
It is shown that the QHR which is defined by the propagator 
in the momentum representation 
exists in finite system if the magnetic field is 
of suitable strength.

(2-a) Disorder potentials

A Hamiltonian, 
$$
\int d\vec x\Psi^\dagger(x)[{(\vec p+e\vec A)^2\over 2m}+V(x)]\Psi(x)
=\sum_{l_1,l_2,R_1,R_2}b_{l_1}(R_1)[E_{l_1}\delta_{l_1,l_2}\delta_{
R_1,R_2}+V_{l_1,l_2}(R_1,R_2)]a_{l_2}(R_2)
\eqno\eq
$$
discribes the systems with a disorder potential $V(x)$.
The kinetic term becomes diagonal and the potential term becomes 
non-diagonal. 
In fact this term becomes a lattice kinetic term if a suitable periodic 
potential is used for $V(x)$, as was shown in Ref(3). 
Eigenstates are extended, then. 
For a short range potential, on the other hand, eigenstates of energy 
$E$ in region, $\vert E-E_l\vert>\delta$ with a small parameter $\delta$, 
becomes localized. 
This is easily seen from the eigenvalue equation,
$$
\sum_{l_2,\vec R_2}
V_{l_1,l_2}(\vec R_1,\vec R_2)u_{l_2}^{(\alpha)}(\vec R_2)=
(E^{(\alpha)}-E_{l_1})u_{l_1}^{(\alpha)}(\vec R_1).
\eqno\eq
$$
Eigenvector $u_{l_1}^{(\alpha)}(m_1,n_1)$ has the same coordinates dependence 
as the transformed potential term $V_{l_1,l_2}(\vec R_1,\vec R_2)$ 
provided $E-E_l\neq0$. 
Hence an eigenvector is localized around a short range impurity if its energy 
$E$ is away from Landau level energy $E_l$. 
We confirmed this property of wave functions by solving the equation 
numerically$^{\c}$. 
Wave functions of the energy $E-E_l\neq0$ have spatial extensions of few 
magnetic lengths. 

Many short range impurity problem is studied as easily as one impurity 
problem if impurities are dilute and random. 
Localized wave functions around one short range impurity decrease 
very fast 
and almost vanishes at nearby impurities if distances between impurities 
are much larger than the magnetic distance. 
Corrections of localized wave functions due to nearby impurities are thus 
very small and is treated perturbatively. 
If $E-E_l$ is also very small, the situation is different and the energy 
denominator becomes very small and compatible to the small numerator in 
perturbative series. 
The correction can become order one, and 
perturbative series may be divergent, then, and wave function becomes 
extended. 
Consequently, one-particle property depends on its energy in the system 
with dilute random impurities. 
Wave function is localized if its energy $E$ is in $\vert E-E_l\vert>\delta$ 
with impurity dependent small parameter $\delta$. 
The propagator in the momentum representation has no singularity in this 
energy region.

(2-b) Boundary potentials

A potential $V_0\theta(x)$ shows a potential barrier in the positive $x$ 
region with a boundary at $x=0$. 
Due to a translational invariance in $y$-direction, it is convenient to 
project the potential and wave functions to those of definite momenta, as 
$$
\eqalign{
V_{l_1,l_2}(m_1,n_1;m_2,n_2)&={a\over 2\pi}\int\nolimits
^{\pi/a}_{-\pi/a}dp_y 
e^{ip_y(n_1-n_2)a}V_{l_1,l_2}(m_1;m_2;p_y),\cr
u_l(m,n)&=\sqrt{a\over 2\pi}\int\nolimits
^{\pi/a}_{-\pi/a}dp_y e^{ip_y na}u_l(m,p_y)
.}
\eqno\eq
$$
The eigenvalue equation becomes, 
$$
\sum_{l_2,m_2}\{E_{l_1}\delta_{l_1,l_2}\delta_{m_1,m_2}
+V_{l_1,l_2}(m_1;m_2;p_y)\}
u^{(\alpha)}_{l_2}(m_2,p_y)=E^{(\alpha)}
u^{(\alpha)}_{l_1}(m_1,p_y).
\eqno\eq
$$
We solved the equation numerically. 
Edge states have continuous energy and wave functions are extended in 
$y$-direction and are confined in $x$-direction.
As the boundary potential gives steep electric field in $x$-direction, 
edge states have current with definite direction and are regarded as 
chiral modes. 
They have been computed in the present representation and 
were given in Figs.7 and 8 of Ref.(3). 
They have continuous energies in the wide range and bridge from 
one Landau level to next Landau level. 
Hence there are always one-dimensionaly extended states around an 
arbitrary value of Fermi energy. 

In systems with finite width in the $x$-direction, there are two edges. 
The wave functions change their properties depending upon 
their widths. 
To see them, we solve systems of potential wells in the $x$-direction 
numerically. 
As are shown in Fig.1 and Fig.2, if the width is large, two edge 
states are separated, but if the width is small, they are combined. 
In wide systems, one-dimensionaly extended states have small overlapp with 
the momentum eigenstates that are used in our method of computing 
the Hall conductance. 
We will see that they give no correction to the quantized Hall conductance 
in certain situation. 

Now we find out when the edge terms and impurities give no correction. 
In the section 5 of the present paper, we apply the momentum representation
of the $\sigma_{xy}$ as in Ref.(1). 
The edge potential term is treated as perturbative term in the 
Hamiltonian and in the propagator. 
In this expansion, infra-red divergence is not involved unless the 
Fermi energy agrees to the Landau level energy. 
Moreover, the series converges well if the energy difference, 
$E-E_l$, is much larger than the perturbative energy of the states 
projected into the momentum eigenstates. 
These momentum eigenstates, although they are fictitious, play the 
important roles in our calculations and a band width of these 
fictitious extended states is important and is 
estimated in the following. 
If the Fermi energy is in the outside of this fictitious band, 
the convergence of the perturbative expansion is very good, and the 
low-energy theorem for the $\sigma_{xy}$ is derived easily. 
It should be noted that this can occur even if the real one-dimensional 
edge states bridge from one Landau level to next Landau level and 
have zero energy around the Fermi energy. 

For estimation of the band width in the momentum representation 
we study eigenvalue equation of the 
Hamiltonian in one method, and we study the propagator in another 
method. 

The band width in the first method is estimated 
directly from Eq.(2.2). 
The boundary potential in magnetic lattice representation 
$V_{l_1,l_2}(\vec R_1,\vec R_2)$ decreases rapidly with the distance 
between the coordinates $\vec R_1$ or $\vec R_2$ and the boundary and 
behave as 
$$
V(\vec R_1,\vec R_2)\sim V_0{\rm Max}
(e^{-{\pi\over 2}({L_1\over a})^2},e^{-{\pi\over 2}({L_2\over a})^2})
,
\eqno\eq
$$
where $V_0$ is the potential height and $L_1(L_2)$ is the distance. 
Hence the equation(2.2) becomes at a point $\vec R$ in the middle of the 
system, 
$$
\vert E^{(\alpha)}-E_{l_1}\vert={\vert \sum_{l_2,\vec R'}
V_{l_1,l_2}(\vec R,\vec R')u_{l_2}^
{(\alpha)}(\vec R')\vert\over\vert u_{l_1}^{(\alpha)}(\vec R)\vert}
\leq{V_0 e^{-{\pi\over 2}({L\over 2a})^2
}\over\vert u_{l}^{(\alpha)}(\vec R)\vert}. 
\eqno\eq
$$
An extended eigenstate which does not vanishes in whole space satisfies, 
$$
\vert u_{l}^{(\alpha)}(\vec R)\vert\geq{c\over\sqrt{L}}.
\eqno\eq
$$
By combining Eq.(2.6) and Eq.(2.7), we have 
$$
\vert E^{(\alpha)}-E_l\vert\leq c V_0 \sqrt{L} 
e^{-{\pi\over 2}({L\over 2a})^2}.
\eqno\eq
$$
The band width, $\Delta_{\rm edge}$, thus satisfies, 
$$
\Delta_{\rm edge}< cV_0\sqrt{L}e^{-{\pi\over 2}({L\over 2a})^2}.
\eqno\eq
$$
The extended states in the momentum representation are generated from 
the extended states which have energies within Eq.(2.8). 

In the second method, we study the propagator. 
As is expressed in Ref.(1), the energy of the momentum eigenstates 
in the system of disorders can be  defined and is expressed from 
diagrams of Fig.3. 
We calculate the lowest non-trivial order contribution here. 
The characteristic properties of the full order are known from them. 
The self-energy correction of the momentum $p$ 
up to $O(V^2)$ is given by
$$
\langle p\vert V\vert p\rangle+
{1\over E-E_l}\sum_{p_i\neq p
}\vert\langle p\vert V\vert p_i\rangle\vert^2,
\eqno\eq
$$
where the state has definite momentum and the boundary potential 
of Eq.(5.5) and impurity potential are used. 
Since the boundary potential is not vanishing only in the finite 
region around the boundary of few magnetic length, the above quantity 
has the magnitude, 
$$
{aV_0\over L}+
{cV_0^2\over E-E_l}({a\over L})^2,
\eqno\eq
$$
with magnetic field independent constant $c$ and a width $L$. 
If the magnetic field is strong enough, or the width $L$ is large 
enough, then the self-energy correction due to the boundary potential 
or due to the boundary potential and the impurity potential 
becomes arbitrary small. 
Hence the width of the fictitious band which appears in our computation 
can be (much) smaller than the Landau level spacing if the $L$ is large 
enough or the magnetic field is strong enough. 
If the Fermi energy is in the outside of this fictitious band, 
the system is in the QHR. 
The QHR can be realized in the finite systems.

(2-c) Electron interactions

Interactions also remove the degeneracy of Landau levels. 
We estimate band width due to Coulomb interactions based on perturbative 
calculation for momentum eigenstates of Fig.4. 
Within the $l$-th Landau level space, the propagator is written as, 
$$
\eqalign{
&{1\over p_0-E_l-\Sigma(\vec p,p_0)},\cr 
&\Sigma(\vec p)=\nu\int{d\vec k\over (2\pi)^2}V(\vec k)\vert (f_l
\vert e^{ik\xi}\vert f_l)\vert^2,
}
\eqno\eq
$$
where $\nu$ is the filling factor in the $l$-th Landau level. 
The eigenvalue of the energy is obtained from the pole of the above 
propagator. 
The energy shift from the degenerate value $E_l$ 
is given by $\Sigma(\vec p)$, 
which vanishes at $\nu=0$, and agrees with 
$$
\Delta E=\int{d\vec k\over (2\pi)^2}V(\vec k)\vert (f_l
\vert e^{ik\xi}\vert f_l)\vert^2=e^2\sqrt{2\pi eB}. 
\eqno\eq
$$
at $\nu=1$ in the lowest Landau level space. 
The difference of the energy correction between $\nu=1$ and $\nu=0$ 
gives the band width due to interactions. 
We have thus
$$
\Delta_{\rm int}=e^2\sqrt{2\pi eB},
\eqno\eq
$$
which has a weaker magnetic field dependence than Landau levels spacing.

The width in our momentum representation due to edge states, Eq.(2.9) 
or Eq.(2.11), 
becomes (much) smaller than the Landau level 
spacing with a strong magnetic field or with a large  $L_x$. 
The width due to interactions, Eq.(2.14), also becomes much 
smaller than the level's spacing. 
Hence with a strong magnetic field, there are wide energy region which 
have no singularity due to two-dimensional extended states. 
The localized states due to short range impurities can have energies in 
these regions. 
We call these regions as quantum Hall regime(QHR). 
QHR exists in systems of boundary, disorder 
and interactions if the magnetic field 
is strong enough. Whole spectrum is shown in Fig.5.

\chapter{Low energy theorem of $\sigma_{xy}$}

Low energy theorem of $\sigma_{xy}$ is satisfied in QHR with impurities 
and interactions. 
We study many-body problems with (quantized) field operators 
$\{a_l(m,n;t),b_l(m,n;t)\}$ which are defined on lattice sites, 
$$
\vec R_{mn}=a(m,n),\ 
a=\sqrt{2\pi\hbar\over eB}
\eqno\eq
$$
and satisfy an equal time commutation relation, 
$$
\{a_{l_1}(m_1,n_1;t_1),b_{l_2}(m_2,n_2;t_2)\}\delta(t_1-t_2)=
\delta_{l_1,l_2}\delta_{m_1,m_2}\delta_{n_1,n_2}\delta(t_1-t_2). 
\eqno\eq
$$
Operator of zero momentum was not included originally in operator set, 
but is added for computational convenience$^{\a}$. 
This operator decouples from physical space due to the constraints of 
von Neumann lattice coherent states. 

The propagator, vertex function, and current correlation functions 
in translationaly invariant systems are defined by,
$$
\eqalign{
&\int dt_1 dt_2\sum_{X_1,X_2}e^{i(p_1x_1-p_2x_2)}\langle T\{a_{l_1}(
\vec X_1,t_1)b_{l_2}(\vec X_2,t_2)\}\rangle={1\over a^2}(2\pi)^3\delta
(p_1-p_2)S_{l_1,l_2}(p_1),\cr
&\int dt_1 dt_2 dx\sum_{X_1,X_2} e^{i(qx+p_1X_1-p_2X_2)}\langle T
\{j_\mu(x)a_{l_1}(\vec X_1,t_1)b_{l_2}(\vec X_2,t_2)\}\rangle=\cr
&\qquad\qquad\qquad
{i\over a^2}(2\pi)^3\delta(p_1-p_2+q)S_{l_1,l'_1}(p_1)
\Gamma_\mu^{l'_1,l'_2}(p_1,p_2)S_{l'_2,l_2}(p_2),\cr
&\int dx_1 dx_2 e^{i(q_1x_1-q_2x_2)}\langle T\{j_{\mu_1}(x_1)
j_{\mu_2}(x_2)\}\rangle=(2\pi)^3\delta(q_1-q_2)\pi_{\mu_1,\mu_2}(q_1).
}
\eqno\eq
$$
$T(\cdot)$ denotes time-ordered product and momenta conjugate to 
lattice coordinates are defined on a torus.  
Hall conductance in QHR is computed from the time ordered current 
correlation functions and agree to that of retarded product used in 
Kubo formula$^{\j}$. 
Owing to the current conservation, 
$$
\partial_\mu j^\mu=0, 
\eqno\eq
$$
and the commutation relations Eq.(3.2) and their representation Eq.(1.9), 
the above Green's functions satisfy$^{\a,\b}$, 
$$
\eqalign{
q^\mu\pi_{\mu\nu}(q)&=\pi_{\mu\nu}(q)q^\nu=0,\cr
q^\mu \Gamma_\mu(p_1,p_2)&=S^{-1}(p_1)R(p_2)-L(p_1)S^{-1}(p_2),\cr
R_{l_1,l_2}(p)&=\delta_{l_1,l_2}+iq_x[d_x(p)\delta_{l_1,l_2}
+\bar d_{x,l_1,l_2}]+iq_y[d_y(p)\delta_{l_1,l_2}
+\bar d_{y,l_1,l_2}],\cr
L_{l_1,l_2}(p)&=\delta_{l_1,l_2}+iq_x[d'_x(p)\delta_{l_1,l_2}
+\bar d'_{x,l_1,l_2}]+iq_y[d'_y(p)\delta_{l_1,l_2}
+\bar d'_{y,l_1,l_2}],
}
\eqno\eq
$$
$$
\eqalign{
&d_i(p)=d'_i(p)=\sum_{X_1-X_2}e^{ip(X_1-X_2)}
\langle{\tilde{\vec X}}_1
\vert (\vec X-\vec X_1)_i\vert \vec X_2\rangle,
\cr
&\bar d_{i l_1 l_2}(p)
=\bar d'_{i l_1 l_2}(p)=(f_{l_1}\vert(\vec\xi)_i\vert f_{l_2}).
\cr
}
\eqno\eq
$$
The explicit forms of $d_i(p)$ and $\bar d_i(p)$ are given in 
Appendix A. 
The Hall conductance is the slope of $\pi_{\mu\nu}(q)$ at the origin 
and is written as
$$
\eqalign{
\sigma_{xy}&={1\over3!}\epsilon^{\mu\nu\rho}{\partial\over\partial q_
\rho}\pi_{\mu\nu}(q)\biggr\vert_{q=0}={e^2\over2\pi}N_w,\cr
N_w&=
{1\over24\pi^2}\int d^3p\epsilon^{\mu\nu\rho}
\Tr[{\partial \tilde S^{-1}(p)\over\partial p_{\rho}}\tilde S(p)
{\partial \tilde S^{-1}(p)\over\partial p_{\mu}}\tilde S(p)
{\partial \tilde S^{-1}(p)\over\partial p_{\nu}}\tilde S(p)].}
\eqno\eq
$$
with the transformed propagator 
$$
\eqalign{
&\tilde S(p)=V(\vec p)S(p)U(\vec p),\cr
&U(p){\partial\over\partial p_i}U^{-1}(p)+l_i(p)=0,\cr
-&{\partial V^{-1}(p)\over\partial p_i}V(p)+r_i(p)=0,\cr
&l_i(p)=r_i(p)=d_i(p)\delta_{l_1,l_2}+\bar d_i(p)\delta_{l_1,l_2}.
}
\eqno\eq
$$
We have used a fact that 
the vertex function which is transformed by the above matrices as 
$$
\tilde \Gamma_\mu(p_1,p_2)=U^{-1}(\vec p_1)\Gamma_\mu(p_1,p_2)
V^{-1}(\vec p_2)
\eqno\eq
$$
satisfies the standard Ward-Takahashi identity, 
$$
\tilde \Gamma_\mu(p,p)={\partial {\tilde S}^{-1}(p)\over\partial p_\mu}
\eqno\eq
$$
It should be noted that the relations Eqs.(3.4)$\sim$(3.10) are 
satisfied in interacting systems as well. 

\ 

The meaning of $N_w$ :

The $N_w$ of Eq.(3.7) is a winding number of mapping from the momentum space 
to a matrix space defined by the propagator $\tilde S_{l_1,l_2}(p)$. 
The momentum in spatial direction is defined on a torus. 
The integration region of the momentum in temporal direction is also 
regarded as a compact space, when the propagator is non-singular at 
$p_0=\pm\infty$. 
Physically, there is no singularity at $p_0=\pm\infty$ in theories we 
study here and magnetic field effect should disappear at $p_0=\pm\infty$. 
Furthermore, the energy integration region becomes a 
closed path in complex plane, which is compact, if the difference 
$N_w(E_{F_1})-N_w(E_{F_2})$ is computed. 
Hence the momentum in temporal direction is also regarded as compact. 

Consequently, the momentum space is regarded as a compact space. 
The matrix space is of infinite dimensions and seems 
to have a difficulty in defining a topological invariant. 
However, since the energy of Landau level $E_l$ diverges if 
$l\rightarrow\infty$, matrix space which corresponds to finite energy 
is of finite dimensional. 
It includes SU(2) space as a subspace. 
Hence $N_w$ generally agrees to an integer as an elements of 
$\pi_3(G),\ G\supset SU(2)$. 

\ 

Computation of $N_w$ :

(i) Free system

We compute $N_w$ in a free system without disorders and interactions 
first. 
Let substitute Eq.(3.6), (3.8), and free propagator 
$S_0(p)$ expressed by $(p_0-E_{l_1}\pm i\epsilon)^{-1}\delta_{l_1,
l_2}$ 
into Eq.(3.7), then we find that the 
integrand of Eq.(3.7) is independent of $\vec p$ and is given by
$$
-{i\over 4\pi^2}{1\over eB}\sum_l{1\over p_0-E_l\pm i\epsilon}.
\eqno\eq
$$
We have, then, 
$$
N_w=l,\quad E_l<E_F<E_{l+1}. 
\eqno\eq
$$
$N_w$ depends on the Fermi energy and agrees to an integer at 
$E\neq E_l$ as is shown in Fig.6. 
The value is ambiguous at $E=E_l$ due to degeneracy of Landau levels. 
By adding momentum dependent small energy, the degeneracy is removed and 
we have a unique value of $N_w$. 
The value thus obtained is proportional to electron filling factor 
$\nu$, as 
$$
N_w=\nu,
\eqno\eq
$$
which leads the following classical value of the Hall conductance:
$$
\sigma_{xy}={e^2\over 2\pi}\nu.
\eqno\eq
$$

(ii) Systems with impurities

Degeneracy of Landau levels are partially removed by impurities and 
there appear one-particle energy 
levels of having energy $E$ at $E\neq E_l$. 
It was shown that the energy eigenstates of $E$ at 
$\vert E-E_l\vert>\delta$, are localized if the impurities are dilute, 
random and short range based on perturbative expansion method in 
von Neumann magnetic lattice representation, where $\delta$ is a 
system dependent small constant. 
Numerical calculations in random system 
also show that the wave functions are localized at $E\neq E_l$ with 
the localization length being inversely proportional to 
some power of $E-E_l$, $(E-E_l)^{-\lambda},\ \lambda>0$. 
Since the wave functions of localized states have finite extensions and 
their energies are discrete, it is possible to compute the contributions 
of the localized states to $\sigma_{xy}$ perturbatively, if the Fermi 
energy is in the range $\vert E_F-E_l\vert>\delta$. 

We assume, hereafter, that the system is in quantum Hall regime where 
all the one-particle states around Fermi energy are localized and 
the momentum conserving propagator in the momentum representation 
has no singularity. 
We show that only the momentum conserving amplitude contributes to 
the linear coefficient of the current correlation function first. 
We decompose the current correlation function into the momentum conserving 
part $\pi^{(1)}_{\mu\nu}(q)$ and the momentum non-conserving part 
$\pi^{(2)}_{\mu\nu}(q_1,q_2)$. 
$$
\pi_{\mu\nu}(q_1,q_2)=(2\pi)^3\delta(q_1-q_2)\pi^{(1)}_{\mu\nu}(q_1)
+\pi^{(2)}_{\mu\nu}(q_1,q_2).
\eqno\eq
$$
In the latter, $\vec q_1$ and $\vec q_2$ are momenta carried 
by currents and are independent each other. 
This term is not zero generally due to a lack of translational 
invariance. 
Current is conserved, hence the identities, 
$$
\eqalign{
&q^\mu\pi^{(1)}_{\mu\nu}(q)=\pi^{(1)}_{\mu\nu}(q)q^\nu=0,\cr
&q_1^\mu\pi^{(2)}_{\mu\nu}(q_1,q_2)=\pi^{(2)}_{\mu\nu}(q_1,q_2)q_2^\nu=0,
}
\eqno\eq
$$
are satisfied. 
In quantum Hall regime, $\pi^{(1)}_{\mu\nu}(q)$ and 
$\pi^{(2)}_{\mu\nu}(q_1,q_2)$ have no singularity at the origin 
$q_\mu=0$ or $q_{1\mu}=q_{2\mu}=0$, because one particle states have 
discrete energies and their wave functions have finite spatial extensions. 
Singularity is not generated. 
Thus they are expanded as, 
$$
\eqalign{
\pi_{\mu\nu}^{(1)}(q)&=\pi_{\mu\nu}^{(1)}(0)+
\pi_{\mu\nu,\rho}^{(1)}(0)q^\rho+O(q^2),\cr
\pi_{\mu\nu}^{(2)}(q_1,q_2)&=\pi_{\mu\nu}^{(2)}(0,0)+
\pi_{\mu\nu;\rho}^{(2)}(0,0)q_1^\rho+
\pi_{\rho;\mu\nu}^{(2)}(0,0)q_2^\rho+O(q^2).
}
\eqno\eq
$$
The coefficients are finite. 
We plug these forms to Eq.(3.16), and we have, 
$$
\eqalign{
&\pi_{\mu\nu}^{(1)}(0)=0,\cr
&\pi_{\mu\nu,\rho}^{(1)}(0)+\pi_{\mu\rho,\nu}^{(1)}(0)=0,\cr
&\pi_{\mu\nu;\rho}^{(1)}(0)+\pi_{\rho,\nu,\mu}^{(1)}(0)=0.
}
\eqno\eq
$$
The slope of $\pi_{\mu\nu}^{(1)}(q)$ at the origin is, hence, totally 
anti-symmetric, and is written with one constant $c$, as 
$$
\pi_{\mu\nu,\rho}^{(1)}(0)=c\epsilon_{\mu\nu\rho}.
\eqno\eq
$$
We have, also, 
$$
\eqalign{
&\pi_{\mu\nu}^{(2)}(0,0)=0,\cr
&\pi_{\mu\nu;\rho}^{(2)}(0,0)=0,\cr
&\pi_{\rho;\mu\nu}^{(2)}(0,0)=0,
}
\eqno\eq
$$
by substituting the second form, Eq.(3.17), into the second equation of 
Eq.(3.16), because $\vec q_1$ and $\vec q_2$ are independent. 
Thus the linear term vanishes if $\vec q_1$ and $\vec q_2$ are 
different and independent. 
In fact, even in systems with translational invariance in one 
direction where one component of $\vec q_i$ are the same, the linear term 
vanishes. 
This is understandable easily from the fact that there is only one free 
parameter left in Eq.(3.16), and one additional condition makes 
non-trivial solution disappear. 
The slope of the current-current correlation function at the origin is due 
to the momentum conserving part, $\pi_{\mu\nu}^{(1)}(q)$, and the momentum 
non-conserving part, $\pi_{\mu\nu}^{(2)}(q_1,q_2)$, does not contribute. 
This will be used also in the next part when interaction effects are 
studied.

(iii) Resummation formula of the $\sigma_{xy}$ in the 
momentum representation. 

We use a representation of the momentum conserving part, 
$\pi_{\mu\nu}^{(1)}(q)$, in terms of a momentum conserving part of 
propagator and a momentum conserving part of vertex part. 
They are defined by summing the momentum conserving parts in 
perturbative expansions. 
The details of the resummation formula have been given in Ref.(1). 
Here, we give only the momentum conserving part of propagator in a 
system of impurity potentials or boundary potentials as an example. 
QHR is defined as the outside region of the energy band of the 
momentum conserving part of the propagator. 
In order to write the propagator 
with a momentum representation in the system of disorder potential $V$, 
we start from, 
$$
\langle p_1\vert{1\over E-(H_0+V)}\vert p_2\rangle=
{1\over E-E_0(p_1)}[\delta_{p_1,p_2}+{\langle p_1\vert V\vert p_2
\rangle\over E-E_0(p_1)}+{\langle p_1\vert V\vert p'
\rangle\over E-E_0(p_1)}{\langle p'\vert V\vert p_2
\rangle\over E-E_0(p')}+\cdots],
\eqno\eq
$$
$$
H_0\vert p\rangle=E_0(p)\vert p\rangle,
$$
where $H_0$ is assumed to be invariant under translations. 
Eigenvalue $E_0(p)$ could have a dependence on the momentum. 
In quantum Hall system, $H_0$ is given by
$$
H_0=\sum E_l b_l^\dagger(R)a_l(R),
\eqno\eq
$$
and the energy eigenvalue, $E_0(p)$, thus is constant, $E_l$. 
Due to the non-invariant term, V, the momentum is not conserved and 
$p_2$ is different from $p_1$ generally. 
The momentum conserving propagator within the $l$-th Landau level 
space is defined by the perturbative series as, 
$$
S(p)=\langle p\vert{1\over E-(H_0+V)}\vert p\rangle
={1\over E-E_l}[1+{\langle p\vert V\vert p
\rangle\over E-E_l}+{\langle p\vert V\vert p'
\rangle\over E-E_l}{\langle p'\vert V\vert p
\rangle\over E-E_l}+\cdots].
\eqno\eq
$$
This can be written in the following form:
$$
\eqalign{
S(p)&
={1\over E-E_l}[1+{\langle p\vert V\vert p
\rangle\over E-E_l}+\sum_{p'=p}{\langle p\vert V\vert p'
\rangle\over E-E_l}{\langle p'\vert V\vert p
\rangle\over E-E_l}+\cdots\cr&
\qquad\qquad+\sum_{p'\neq p}{\langle p\vert V\vert p'
\rangle\over E-E_l}{\langle p'\vert V\vert p
\rangle\over E-E_l}+\cdots]\cr&
={1\over E-E_l}[1+{\langle p\vert V\vert p
\rangle\over E-E_l}+({\langle p\vert V\vert p
\rangle\over E-E_l})^2+({\langle p\vert V\vert p
\rangle\over E-E_l})^3+\cdots\cr&
\qquad\qquad+\sum_{p'\neq p}{\langle p\vert V\vert p'
\rangle\over E-E_l}{\langle p'\vert V\vert p
\rangle\over E-E_l}+\cdots]
={1\over E-E_l-\Sigma(E,p)},
}
\eqno\eq
$$
$$
\Sigma(E,p)={\langle p\vert V\vert p\rangle}+
\sum_{p'\neq p}{\langle p\vert V\vert p'\rangle 
\langle p'\vert V\vert p\rangle\over E-E_l}+\cdots.
\eqno\eq
$$
The above formula is equivalent to write the full Green's function 
in interacting systems with a self-energy part and a free part. 
The self-energy part is defined from the one-particle irreducible 
part in standard manner. 
In the present system, the one-particle irreducible part is defined 
from the momentum, and the self-energy part $\Sigma(E,p)$ 
is defined based on it. 
It is easy to see it in diagrams. 
They are given in Fig.3. 

In the above perturbative expansion, if the energy denominator 
$E-E_l$ does not vanish, there is no infra-red divergence. 
Moreover the series converge fast if each term in $\Sigma(E,p)$ 
is small. 
In fact, the boundary potential $V$ has a non-zero value only in 
narrow regions near the boundaries. 
Hence the matrix element from the boundary potential, 
$\langle p\vert V\vert p'\rangle$ is of order $V_0 a/L$. 
There are the following $\Delta_1$ and $\Delta_2$ : 

\noindent
For such $E$ that satisfies
$$
\vert E-E_l\vert>\Delta_1,
\eqno\eq
$$
$\Sigma(E,p)$ satisfies,
$$
\Sigma(E,p)<\Delta_2,
\eqno\eq
$$
$$
\Delta_2<\Delta_1,
$$
where $\Delta_1$ is of order $V_0 a/L$. 
The convergence of Eq.(3.25) is good then, and $S(p)$ has no singularity. 
Hence the energy region of Eq.(3.26) is regarded as QHR. 
They have the width of order $a/L$ and becomes very small if $L$ 
is large or the magnetic field is strong. 
It should be noticed that $\Delta_1$ is finite and QHR exists 
if $\Delta_1$ is smaller than Landau level energy spacing 
$E_{l+1}-E_l$. 
This is possible even in a finite $L$ and a finite magnetic field case. 

$S_0(p)$ in free systems are replaced with $S(p)$ in systems 
with disorders. 
In QHR, $S(p)$ has no singularities and the perturbative expansions 
due to disorders which modifies $S_0(p)$ into $S(p)$, 
converge well. 
Their effects in the $\sigma_{xy}$ are also studied by perturbative 
expansions and they do not give any corrections to 
the quantized Hall conductance from the Coleman-Hill argument. 
Consequently the quantization of the $\sigma_{xy}$ is exact in QHR 
that is defined by $S(p)$ of Eq.(3.23). 
The fact that the QHR is defined by the momentum representation, 
$S(p)$, becomes important when we discuss the edge states 
generated by the boundary potential.

(iv) Systems with interactions in 
quantum Hall regime, $\vert E_F-E_l\vert>\delta$. 

Since all the one-particle states are localized in the energy region, 
$\vert E-E_l\vert>\delta$, current correlation function has no 
singularity if the Fermi energy is in this energy region. 
Since $\delta$ and $\Delta_1$ should be the same order, we identify them. 
We investigate this energy region here and show that the interactions 
do not give any corrections to $\sigma_{xy}$ in QHR. 

As was shown in the previous part, the momentum non-conserving part of 
the current correlation function has no first derivative 
at the origin and does not 
contribute to $\sigma_{xy}$. 
A momentum conserving part in a system with interactions 
also has no first derivative at the origin if that agrees to 
a (regular) limit of a momentum non-conserving part. 

In QHR, $\pi_{\mu\nu}(q)$ has no singularity in $q_\mu$. 
Perturbative higher order diagrams due to interactions and disorders 
also have no singularity. 
They are computed by integrating a product of propagators and vertices. 
By cutting one of bosonic propagators and regarding them as two external 
lines with different momenta, following Coleman and Hill$^{\k}$, 
we are able to regard the original 
amplitude as a limit of momentum non-conserving amplitudes, 
$\tilde \pi^{(2)}_{\mu\nu}(q_1,q_2)$, which satisfy the same identity 
as $\pi_{\mu\nu}^{(2)}(q_1,q_2)$ of Eq.(3.14). 
Hence, $\tilde \pi_{\mu\nu}(q_1,q_2)$ has no linear term in $q^{(i)}_\mu$ 
and $\tilde \pi_{\mu\nu}(q_1,q_2)\vert_{q_2\rightarrow q_1}$ 
does not contribute to $\sigma_{xy}$ in QHR, as well. 

Only the momentum conserving part that is isolated from the momentum 
non-conserving parts and is not connected 
with any non-conserving parts contributes to $\sigma_{xy}$. 
There are two kind of momentum conserving parts that are isolated from 
non-conserving parts. 
The first type is the lowest order diagram, i.e., one loop diagram. 
This is obviously isolated from any other diagrams. 
The second type is similar to the first one, but is a loop diagram 
with a dressed momentum dependent propagator. 
The momentum dependent propagator expresses extended states and has a 
singularity within a finite band width, which is estimated later. 
If the Fermi energy is in the outside of the continuous energy band 
region, 
all the particle states around the Fermi energy are localized and the 
system is regarded as QHR.

Thus the higher order terms do not contribute to $\sigma_{xy}$ in QHR. 
If there are ultra-violet divergences, physical observables are 
computed with renormalized quantities. 
The bare charge is replaced with the renormalized charge. 
Consequently, $\sigma_{xy}$ is given by exactly quantized value 
$(e^2/h)N$ in QHR with the renormalized charge $e$. 

\chapter{Cylider geometry}

We study electron system of a cylinder geometry with a length $L$. 
For simplicity we assume $L$ is an integer multiple of lattice spacing 
$a$. 
Wave functions in center coordinates $\vec R_{mn}$ satisfy, 
$$
\Psi(\vec R_{m,n}+L \vec e_x)=\Psi(\vec R_{m,n}),
\eqno\eq
$$
and the corresponding discrete momentum $p_x$ satisfies,
$$
L p_x=2\pi n,\ n={\rm integer}.
\eqno\eq
$$
If the $L$ is much larger than magnetic lattice spacing $a$, 
localized one particle states are essentially the same as those of the 
previous infinite system, 
because they have finite spatial extensions and 
are insensitive to the boundary conditions. 
Hence one-particle states in the energy region $\vert E-E_l\vert >
\tilde \delta$ with a slightly modified $\tilde \delta$ are localized. 
This energy region corresponds to QHR. 
We study a finite size effect of $\sigma_{xy}$ in QHR. 

The Hall conductance in QHR of a cylinder geometry is given by, 
$$
\sigma_{xy}={e^2\over 2\pi}\sum_{p_x}\int dp^0 dp_y\epsilon_{\mu\nu\rho}
{1\over 24\pi^2}\Tr [{\partial \tilde S^{-1}\over\partial p_\mu}\tilde S
{\partial \tilde S^{-1}\over\partial p_\nu}\tilde S
{\partial \tilde S^{-1}\over\partial p_\rho}\tilde S],
\eqno\eq
$$
where the $p_x$ integration in Eq.(3.7) was replaced with the discrete 
momentum summation. 
By the replacement, the integral is generally changed depending upon the 
form of the integrand. 
In the present case, however, the integrand becomes, 
$$
\epsilon_{\mu\nu\rho}
{1\over 24\pi^2}\Tr [{\partial \tilde S^{-1}\over\partial p_\mu}\tilde S
{\partial \tilde S^{-1}\over\partial p_\nu}\tilde S
{\partial \tilde S^{-1}\over\partial p_\rho}\tilde S]=
-{i\over 4\pi^2}{1\over eB}\sum_l {1\over p_0-E_l\pm i\epsilon},
\eqno\eq
$$
where $-i\epsilon(+i\epsilon)$ is taken if the energy $E_l$ is less 
than(larger than) Fermi energy, 
and does not depend on $p_x$. 
Hence $\sigma_{xy}$ is unchanged. 
$\sigma_{xy}$ in the cylinder geometry is identical to that of the 
infinite system. 
There is no finite size correction in quantized Hall conductance in QHR. 

The effects of impurities and interactions in QHR are studied in the 
same way as infinite systems. 
They do not give any corrections. 
We will describe this in more detail in the next section. 

\chapter{Hall bar geometry}

We study realistic systems which have Hall bar geometry in this section. 

(5-1) finite system

Realistic experiments are done with electronic systems of Hall bar 
geometry. 
In one direction a Hall bar system has potential barriers which confine 
electrons in inside. 
There are gates in perpendicular direction through which electrons move 
in and move out. 
So it would be sufficient to study a system that is finite in one 
direction due to potential 
barrier and infinite in another direction. 

The potential barrier in the positive $x$ region is expressed with the 
following potential term : 
$$
\eqalign{
&H_I=\int d\vec x V(x)\Psi^\dagger(x)\Psi(x),\cr
&V(x)=V_0(\theta(-x)+\theta(x-L)). 
}
\eqno\eq
$$
The potential is invariant under translation in the $y$-direction, hence 
the eigenstates are the plane wave in this direction. 
The eigenvalue equations were solved before in the present representation. 
The edge states which have the $p_y$ dependent continuous energy in the 
region $E_l<E<E_l+V_0$ and are localized in the $x$-direction are found. 
They cross the Fermi energy. 
Since edge states have the continuous energy around the Fermi energy, 
they could contribute to conductance, generally. 
However, they are confined along the edges and move in one direction 
without reflected. 
We see that they do not contribute to the conductance, then. 

The one-dimensional edge states near the Fermi energy are chiral and 
electrons move in one direction. 
Their properties are unchanged by adding disorder potentials, because 
in the wave equation for a chiral mode with a potential term,
$$
[i{\partial\over\partial x}+V(x)]\Psi=E\Psi,
\eqno\eq
$$
the potential can be removed completely by a gauge transformation
$^{\l}$, 
$$
\eqalign{
&\Psi=e^{i\phi(x)}\tilde\Psi,\cr
&i{\partial\over\partial x}\tilde\Psi=E\tilde\Psi.
}
\eqno\eq
$$
Hence the phase factor of chiral modes are modified by the potential 
but the waves are neither reflected nor localized. 
They carry electric current but do not contribute to conductance 
in the above situation. 

We use the resummation formula in the momentum representation 
and study the finite size effect. 
If one edge state at one side interacts with one edge 
state at another side, 
extended states can be formed, then the propagator in the momeutum 
representation has singularities. 
The effect of the edge states and the boundary potentials are found 
by writing the Hamiltonian as,
$$
H=\sum b_{l_1}(m_1,n_1)a_{l_2}(m_2,n_2)\{E_{l_1}\delta_{l_1,l_2}
\delta_{m_1,m_2}\delta_{n_1,n_2}+V_{l_1,l_2}(m_1,n_1;m_2,n_2)\}.
\eqno\eq
$$
The transformed potential $V_{l_1,l_2}(m_1,n_1;m_2,n_2)$ vanishes in 
inside region and becomes diagonal form $V_0\delta_{l_1,l_2}
\delta_{m_1,m_2}\delta_{n_1,n_2}$ in outside region. 
In narrow boundary region of few magnetic lengths, it has off diagonal 
term. 
The potential and eigenfunctions are transformed in the $y$-direction 
as Eq.(2.3), and satisfy Eq.(2.4). 
We solved the equations numerically. 
Energy eigenvalues and corresponding eigenfunctions are given in Fig.2. 
Ovbiously, edge states are localized in narrow region in  $x$-direction 
and have energy in a range $E_l<E<E_l+V_0$. 
This is understandable easily from the properties of 
$V_{l_1,l_2}(m_1,n_1;m_2,n_2)$. 

Edge state at one boundary could couple with edge state at another 
boundary if their distance is small and there are impurities 
and interactions. 
They could give singularities in the momentum representation, 
$S(p)$, then, and contribute to the 
conductance. 
To study their effects, it is convenient to 
write the previous Hamiltonian, Eq.(5.4), 
in the following manner : 
$$
\eqalign{
&H=H_0+H_1+H_2, \cr
&H_0=\sum_{\vec X\ \rm inside}E_l b_l(\vec X_1) a_l(\vec X_1),\cr
&H_1=\sum_{\vec X\ \rm outside}(E_l+V_0)b_l(\vec X) a_l(\vec X),\cr
&H_2=\sum_{\rm boundary}(\delta V)_{l_1,l_2}(\vec X_1,\vec X_2)
b_{l_1}(\vec X_1)a_{l_2}(\vec X_2).
}
\eqno\eq
$$
Single particle energy from $H_0$ and $H_1$ are $E_l$ or $E_l+V_0$ 
respectively. 
$(\delta V)_{l_1,l_2}(\vec X_1,\vec X_2)$ is not vanishing only when 
$\vec X_1$ and $\vec X_2$ are in the narrow boundary regions 
of few magnetic distances, 
hence a perturbative treatment of $H_2$ is possible. 
When the Fermi energy is slightly larger than $E_l$ but is much smaller 
than $E_l+V_0$, the one particle eigenstates of $H_1$ decouple and 
contribute to conductance only through virtual effects. 
Hence we take into account $H_0$ and $H_2$ meanwhile. 
$H_2$ is taken perturbatively and virtual effects are studied later. 
We study particles confined in the finite inside region. 

We express the field operators of the Landau level in momentum 
representation as, 
$$
\eqalign{
a_l(\vec X)&=\sum_{p_x}\int\nolimits^{\pi/a}\nolimits_{-\pi/a}dp_y
{1\over a}
e^{i\vec p\vec X}a_l(\vec p),\cr
b_l(\vec X)&=\sum \int\nolimits^{\pi/a}\nolimits_{-\pi/a}dp_y
{1\over a}
e^{i\vec p\vec X}b_l(\vec p),\cr
p_x&={2\pi\over L_x}n_x+\alpha.
}
\eqno\eq
$$
In Eq.(5.6), $n_x$ is an integer and a parameter $\alpha$ depends on 
boundary condition. 
The current operator, $\tilde J_\mu$, which is expressed with these 
operators, is conserved within the Hilbert space of confined particles. 
It satisfies, 
$$
\eqalign{
&\partial^\mu \tilde J_\mu=C,\cr
C=&\sum_{\vec X\ {\rm or}\ \vec Y\ \rm in\ outside}b(X)\Gamma a(Y),\cr
\langle {\rm Confined\ particle}&\vert C\vert {\rm Confined\ particle}
\rangle=0.
}
\eqno\eq
$$
The current operators, $\tilde J_\mu$, satisfies the same commutation 
relation as that of infinite system, Eq.(3.4). 
Hence the identities of the infinite system, Eq.(3.5), are satisfied, 
but slight modifications are necessary. 
Current correlation function is expanded with the momentum carried by 
current, which becomes discrete, and its slope at the origin, 
which is simple derivative, is 
proportional to the conductance. 
One loop diagram contributes in QHR, from Coleman-Hill theorem. 
The momentum has a discrete component, from Eq.(5.8), hence the Hall
conductance agrees to that of the torus geometry, Eq.(4.3), which 
has no finite size correction and is independent of $L_x$. 

\ 
{\it Coleman-Hill theorem in finite systems. }

In the present geometry the $x$-coordinate is defined in a finite region 
and the corresponding momentum, $p_x$ becomes discrete. 
Hence the derivations of Ward-Takahashi identity and related 
low energy theorems should be re-examined. 

We define the current correlation function, vertex function, and the 
propagator as Eq.(3.3). 
For a convenience, we use a notation 
$j_\mu(x)$ instead of $\tilde j_\mu(x)$ of 
the previous part. 
We concentrate to the cases where Green's functions 
in configulation space vanish 
in the outside region due to the finite Fermi energy that is 
smaller than $V_0$. 
Integrations of the $x$ coordinate, then, are defined in a 
finite region, $0\leq x\leq L$, and the corresponding momenta thus become
discrete from completeness. 

The current expectation value is connected with external vector potential 
through the current correlation function as 
$$
\eqalign{
&J_\mu(x)=\int dy\pi_{\mu\nu}(x,y)A_\nu(y),\cr
&\pi_{\mu\nu}(p_1,p_2)=\int dx_1 dx_2 e^{ip_1x_1+ip_2x_2}
\pi_{\mu\nu}(x_1,x_2).
}
\eqno\eq
$$
The current correlation function has a translational invariant term and a 
non-invariant term. 
They are expressed by 
$$
\eqalign{
&\pi_{\mu\nu}(p_1,p_2)=\pi^{(1)}_{\mu\nu}(p_1,p_2)+
\pi^{(2)}_{\mu\nu}(p_1,p_2),\cr
&\pi^{(1)}_{\mu\nu}(p_1,p_2)=(2\pi)^2\delta^{(2)}(p_1+p_2)L\delta^{(1)}
_{p_1^x+p_2^x,0}\pi_{\mu\nu}^{(1)}(p_1),\cr
&\pi^{(2)}_{\mu\nu}(p_1,p_2)=(2\pi)\delta^{(1)}(p_1^0+p_2^0)
\tilde\pi_{\mu\nu}^{(2)}(\vec p_1,\vec p_2;p_1^0),\cr
{\rm\ or\ }&\pi^{(2)}_{\mu\nu}(p_1,p_2)=
(2\pi)^2\delta^{(1)}(p_1^0+p_2^0)\delta^{(1)}(p_1^y+p_2^y)
\tilde\pi_{\mu\nu}^{(2)}(p_1^x,p_2^x),
}
\eqno\eq
$$
where $\delta(x)$ is Dirac delta function and $\delta_{m,0}$ is Kronecker 
delta. 
If the Fermi energy is in the energy gap region or in the localized 
state region where all the energy eigenstates around the Fermi energy 
are localized, $\pi_{\mu\nu}^{(1)}(p)$ and 
$\tilde\pi^{(2)}_{\mu\nu}(p_1,p_2)$ 
have no singularities. 
They satisfy, 
$$
\eqalign{
&p^\mu\pi_{\mu\nu}^{(1)}(p)=p^\nu\pi_{\mu\nu}^{(1)}(p)=0,\cr
&p_1^0\tilde \pi^{(2)}_{0\nu}+p_1^i \tilde\pi^{(2)}_{i\nu}=
p_2^0\tilde\pi^{(2)}_{\mu0}(p_1,p_2)+p_2^i\tilde\pi^{(2)}_{\mu i}=0,
}
\eqno\eq
$$
and are expanded with momenta as
$$
\eqalign{
&\pi^{(1)}_{\mu\nu}(p_1)=C\epsilon_{\mu\nu\rho}p_1^\rho+{\rm higher\ 
power},
\cr
&\pi^{(2)}_{\mu\nu}(p_1,p_2)=O(p_i^2),
}
\eqno\eq
$$
from the arguments of Section 3. 
Note that the coefficient $C$ and higher power terms are defined 
uniquely by the slope or the higher order curvaures at the origin 
and are defined by the simple derivatives, even though the momentum is 
not infinitesimal quantity in finite systems. 
We see that $C$ contributes to the Hall conductance, because the total 
current is given by, 
$$
\int dx_0 dx'_2\int\nolimits^L_0 dx'_1J_y(x'_1)=
\int{dp^0_2dp_2^y\over(2\pi)^2}{1\over L}\sum_{p_2^x}\{
\pi^{(1)}_{y,\nu}(0,p_2)+\pi^{(2)}_{y,\nu}(0,p_2)\}A_\nu(p_2).
\eqno\eq
$$
At a point of measurement there is no electric field, 
because that is a gate, and derivative 
of $A_\nu$ vanishes. 
Hence, 
$\pi^{(2)}_{\mu\nu}(p_1,p_2)$ and higher terms of $\pi^{(1)}_{\mu\nu}(p)$ 
thus do not contribute to the Hall conductance. 

Higher order correction due to interactions in quantum Hall regime does 
not exist in $C$, as in the infinite system. 
Higher order corrections of $\pi^{(1)}_{\mu\nu}(p)$ come from diagrams 
in which all the internal momenta satisfy boundary condition. 
If an internal momentum does not satisfy boundary condition
of Eq.(5.6), the momentum 
in that direction is not conserved. 
Consequently, this kind of higher order diagram contributes to only 
$\pi^{(2)}_{\mu\nu}(p_1,p_2)$, and does not contribute to $C$. 
Examples are given in Fig.7. 

As was done by Coleman and Hill, in order to study 
$\pi_{\mu\nu}^{(1)}(p)$ 
we study new diagrams in which internal 
lines of $\pi_{\mu\nu}^{(1)}(p)$ 
are cut and two different momenta, $l_1$ and $l_2$, are given to 
these lines. 
Whenever $l_1+l_2\neq0$, two current have different momenta. 
These diagrams then correspond to $\pi^{(2)}_{\mu\nu}(p_1,p_2)$, 
and are 
written as $\pi^{(2)}_{\mu\nu}(p_1,p_2)=\delta^{(3)}(p_1+p_2+l_1+l_2)
\tilde\pi_{\mu\nu}(p_1,p_2)$, 
and do not contribute to $C$ from the above argument. 
Amplitude $\tilde \pi_{\mu\nu}(p_1,p_2)$ have no singularity in QHR 
and linear terms in $p_1$ or $p_2$ are not allowed, 
because $p_1$ and $p_2$ are independent. 
Then, $\tilde \pi_{\mu\nu}(p,p)$ has no linear term also. 
Thus, none of higher order diagrams contribute to $C$ and 
only the one loop diagram contributes to $C$. 

The linear coefficient is written as a topological invariant. 
Since the current conservation and the commutation relations are the 
same as the infinite system, 
the transformed vertex part as Eq.(3.8) and (3.9), hence satisfies, 
$$
q^\mu\tilde\Gamma_\mu(p_1,p_2)=\tilde S^{-1}(p_1)-\tilde S^{-1}(p_1+q).
\eqno\eq
$$
Consequently, the derivative ${\partial \tilde 
S^{-1}\over\partial p_\mu}$ is 
obtained by comparing a linear coefficient of both sides as
$$
{\partial \tilde S^{-1}(p)\over\partial p_\mu}=
\tilde\Gamma_\mu(p,p).
\eqno\eq
$$
The above equation is satisfied with the simple derivative in the left 
hand side, 
despite of the fact that one component 
of the momentum is discrete. 
Using this formula, the linear coefficient of the current correlation 
function is expressed as,
$$
\sigma_{xy}=
{e^2\over3!}\sum\epsilon^{\mu\nu\rho}{\partial\over\partial p_\rho}
\pi_{\mu\nu}(p)\biggr\vert_{p=0}={e^2\over 2\pi}{1\over 24\pi^2}
\sum_{q_x}
\int d^2 q \epsilon_{\mu\nu\rho}\Tr [
{\partial \tilde S^{-1}\over\partial p_\mu}\tilde S
{\partial \tilde S^{-1}\over\partial p_\nu}\tilde S
{\partial \tilde S^{-1}\over\partial p_\rho}\tilde S].
\eqno\eq
$$
This formula gives topologically invariant expression of the Hall 
conductance in finite systems. 

\ 

(5-2) Boundary effect

We study off-diagonal term $H_2$ of Eq.(5.5). 
This term does not conserve momentum in the $x$-direction. 
Hence perturbative expansion of this term gives higher order diagrams 
in which the total momuntum is not conserved in the 
$x$-direction in addition to the momentum conserving diagrams. 
These higher order diagrams do not contribute to the 
$\sigma_{xy}$ as far as they are treated perturbatively. 
Perturbative treatment is good if the Fermi energy is in the outside of 
the continuous band of the momentum conserving 
propagator, but may not be good if 
the Fermi energy is in the inside of fictitious band of section 2. 
Real extended states could be formed by boundary potentials in a finite 
energy region around the center of Landau level. 
The width of the fictitious band in our momentum representation 
can be made small  in certain conditions discussed before. 
The QHR is in the outside region of the energy bands of
real extended states and of fictitious extended states. 
The width of the bands was estimated in Section 2 and was 
given by Eqs.(2.9) and (2.11). 
Suitably large $L_x$ or strong magnetic field make the widths 
much smaller 
than the Landau level energy spacing, $E_l-E_{l-1}=eB/m$. 
There are enough spacings for the localized states, then, and QHR do 
exist in this situation. 

In QHR, the momentum non-conserving term gives no contribution to the 
slope of current correlation function from the arguments of 
Eqs.(5.11) and (5.12). 
$H_2$ in Eq.(5.5) does not conserve momentum in the $x$-direction but 
conserves in the $y$-direction.  
Consequently, impurities and interactions do not modify the value of 
the quantized Hall conductance there. 

In the inside of extended states energy band
of $S(p)$, the expression Eq.(5.15) 
is valid but there are corrections 
to the quantized $\sigma_{xy}$ and to $\sigma_{xx}$. 
Their contributions were found before, and are given by
$$
\eqalign{
&\sigma_{xy}={e^2\over 2\pi}N+\epsilon,\cr
&\sigma_{xx}=\epsilon',\cr
&\epsilon'\approx\epsilon,
}
\eqno\eq
$$
where $\epsilon$ and $\epsilon'$ are small parameters that are 
proportional to the number of the extended states. 

\ 

(5-3) Virtual effects(renormalization effect)

Higher energy states, such as higher Landau level states, and the states 
in the outside region expressed in $H_1$, give only virtual effect 
in higher order corrections. 
In the infinite system, by using the Ward-Takahashi identity derived 
from the current conservation, the charge 
renormalization factor cancels exactly with the vector potential 
renormalization factor$^\a$ 
and the final formula of the Hall conductance, 
$$
\eqalign{
&\sigma_{xy}={e^2\over h}N_w,\cr
&N_w={1\over 24\pi^2}\int dq\epsilon_{\mu\nu\rho}\Tr[
{\partial \tilde S^{-1}(p)\over\partial p_\mu}\tilde S(p)
{\partial \tilde S^{-1}(p)\over\partial p_\nu}\tilde S(p)
{\partial \tilde S^{-1}(p)\over\partial p_\rho}\tilde S(p)]
}
\eqno\eq
$$
was given. 
Now in a finite system, the current conservation and the 
Ward-Takahashi identity are satisfied. 
Hence we have the formula, 
$$
\eqalign{
&\sigma_{xy}={e^2\over h} N'_w,\cr
&N'_w={1\over 24\pi^2}\int d^2q\sum_i \epsilon_{\mu\nu\rho}\Tr[
{\partial \tilde S^{-1}\over\partial p_\mu}\tilde S
{\partial \tilde S^{-1}\over\partial p_\nu}\tilde S
{\partial \tilde S^{-1}\over\partial p_\rho}\tilde S].
}
\eqno\eq
$$
By combining Eq.(4.4) and Eq.(5.18), we have the Hall 
conductance in a QHR of finite system as, 
$$
\sigma_{xy}={e^2\over h}N,\ N={\rm integer}
\eqno\eq
$$
The $\sigma_{xy}$ has no finite size corrections in QHR.

(5-4) Finite current effects.

The linear response formula, Eq.(5.8), is valid actually if the current 
and the vector potential are infinitesimal,
$$
\delta J_\mu(x)=\int dy\pi_{\mu\nu}(x,y)\delta A_\nu(y). 
\eqno\eq
$$
In order to know the relation for a finite current, we integrate 
the above relation. 
That is possible once the current correlation function is computed in 
a system of a finite vector potential. 
Our formula and theorem can be applied to such correlation function. 
Then the infinitesimal current, infinitesimal potential, and 
current correlation function depend inplicitly on the vector potential. 
We parametrize them with a parameter $s$, under boundary conditions 
$$
\eqalign{
&J_\mu(x,0)=0,\ J_\mu(x,1)=J_\mu(x),\cr
&A_\mu(x,0)=0,\ A_\mu(x,1)=A_\mu(x),\cr
&\delta J_\mu(x,s)=\int dy\pi_{\mu\nu}(x,y;A_\mu(s)
)\delta A_\nu(y,s).
}
\eqno\eq
$$
A total current satisfies a similar equation, 
$$
\eqalign{
&\delta I_x(s)=\sigma_{xy}[A_\mu(s)]\delta V_y(s),\cr
&V_y(0)=0,\ V_y(1)=V_y,\cr
&I_x(0)=0,\ I_x(1)=I_x.
}
\eqno\eq
$$
Now, the $\sigma_{xy}$ could have the same value during the change from 
$s=0$ to $s=1$, if the electronic system is in the same QHR. 
Then the relation for a finite current is obtained by integrating 
Eq.(5.22) with $s$, and becomes 
$$
I_x=\sigma_{xy}V_y,\ \sigma_{xy}={e^2\over h}N.
\eqno\eq
$$

Note that this is possible only in the plateau region$^{\m}$. 
On the other hand, if the system moves from one QHR to extended states 
region or to another QHR, then we are not sure if linear relation 
between a finite current and a finite voltage is satisfied. 
The relation may become complicated, then. 
We will give more arguments of finite current effects, especially 
on a breakdown of the QHE, in the next section. 

The relation between the finite total current and the finite total 
voltage becomes linear in the plateau regions, and the conductance 
$\sigma_{xy}$ is quantized. 
It behaves as Fig.8.

\chapter{Current distribution, B\"uttiker-Landauer formula, 
and Breakdown of QHE.}

In our proof of the quantum Hall effect given in the previous parts, 
current distribution is irrelevant. 
The Hall conductance is the ratio between the total current in one 
direction and total voltage in another direction and so is 
quantized exactly in QHR. 
The current density is not uniform and varies with spatial region, 
generally. 
We study its implication in this section.

There is a completely different approach 
of the quantum Hall effect from ours. 
In its proof, it is assumed that edge states are the only current 
carring states around the Fermi energy 
and is used B\"uttiker-Landauer formula for one-dimensional 
systems. 
The formula may be valid only under the assumption. 
The current distribution is thus important. 
Two approaches give totaly different result when the current 
becomes larger. 
We study the current distribution and finite current effect 
in this section. 
We will see that in general situation, 
the bulk states as well as the edge states carry the current. 
Hall electric field is thus generated at the bulk and it 
leads Landau level broadening. 
QHR becomes narrow consequently and vanishes eventually. 
Breakdown of QHE occurs. 

\ 

\noindent
(6-1) Current distribution

In order to compute the current distribution, we compute an effective 
action of an external vector potential. 
The vector potential is regarded as either the external potential 
added to the system or as a Lagrange multi-plier which is expressing 
a condition of a finite external current. 
Variational principle to the effective action gives the current 
distribution. 

We study the system in which the 
vector potential couples with electron field in a gauge invariant 
manner as, 
$$
\eqalign{
&\int dx[\Psi^\dagger(i\hbar{\partial\over\partial t}+eA_0)\Psi-
\Psi^\dagger{(\vec p+e\vec A_0+e\vec A)^2\over 2m}\Psi-\Psi^\dagger
\Psi V_b(x)],\cr
&\vert\vec \nabla\times\vec A_0\vert=B,\cr
&V_b:{\ \rm boundary\ potential}.
}
\eqno\eq
$$
We integrate the electron fields $\Psi(x)$ and $\Psi^\dagger(x)$ and 
obtain the effective action of $A_0$ and $\vec A$. 
Long distance phenomena are represented by low dimensional terms of the 
effective action$^{\n}$,
$$
\eqalign{
&{1\over2}\int dx[c(x)F^2_{0x}+e(x)F^2_{0y}-d(x)F^2_{ij}]+
{\sigma_{xy}\over 2}\int dx\epsilon_{\mu\nu\rho}A^\mu\partial^\nu
A^\rho+\cdots, \cr
&F_{0x}=\partial_0 A_x-\partial_x A_0,\ 
F_{0y}=\partial_0 A_y-\partial_y A_0,\ 
F_{xy}=\partial_x A_y-\partial_y A_x,\ 
}
\eqno\eq
$$
The $x$-integration is defined in a finite region from $0$ to $L$. 
The coefficients $c(x),\ e(x)$, and $d(x)$ of the above action are 
the constants in the bulk and change their values near the edge. 
They are computed as, 
$$
\eqalign{
&c(x)=e(x)={e^2\over4\pi}{m\over eB},
\cr
&d(x)={3\over 4\pi}{e^2\over m},
}
\eqno\eq
$$
in the bulk. 
Near the edges they change the values. 
Because we need only $c(x)$ here, we give $c(x)$ 
near the edges for a boundary 
potential with a constant slope, $E_b$. 
$$
c(x)={e^2\over 8\pi}{m\over eB}-e^2{B\over4E_b}\sqrt{2\over eB\pi}
e^{-2eB\Delta x^2},
\eqno\eq
$$
where $\Delta x$ is the distance between the boundary and the 
coordinate $x$. 
The first term comes from inter Landau levels and the second term 
comes from intra Landau levels. 
They have opposite sign. 
For high potential barrier, the first term is dominant and 
$c(x)$ is positive definite in all the regions. 
We study this case here and discuss general cases in 
Appendix C.

The total current in the $x$-direction is given by, 
$$
I_y=\int\nolimits^{L}_{0}dx J_y=\sigma_{xy}\{A^0(L)-A^0(0)\}.
\eqno\eq
$$
We assume the translational invariance in the $y$-direction and 
obtain a stationary solution of the effective action under the constraint 
of total current $I_y$ with time independent ansatz :
$$
\partial_0 A_i=0,\ 
\partial_y A_0=0.
\eqno\eq
$$
Then we minimize the following action, 
$$
{1\over2}\int 
d\vec x[c(x)(\partial_x A_0)^2]-\mu\sigma_{xy}\int\nolimits
_{0}^{L}dx
\partial_x A_0(y_2). 
\eqno\eq
$$
Euler-Lagrange equation is given by, 
$$
\partial_x(c(x)\partial_x A_0)=0,
\eqno\eq
$$
and is solved as
$$
\eqalign{
&c(x)\partial_x A_0={\rm const}=C_0,\cr
&\partial_x A_0={C_0\over c(x)},\ c(x)\neq 0. 
}
\eqno\eq
$$
The constant $C_0$ is given from the condition of total current, 
$$
C_0\int\nolimits
_{x_1}^{x_2}dy{1\over c(x)}={I_y\over\sigma_{xy}}={\rm const}.
\eqno\eq
$$
Obviously, if the coefficient $c(x)$ were constant, the local electric 
field, as well as the local current density, would be uniform. 
On the other hand, if $c(x)$ varies, electric field varies also. 
In fact from Eq.(6.4), 
$c(x)$ decreases toward the edge and the electric 
field and current density increases toward the edge. 
The edge current is only a portion of the total current. 
In an exceptional case, where $c(x)$ vanishes or becomes negative 
at the edge and stays at the constant value in the bulk, 
there is an energy minimizing solution which has a 
current only at the edge region. 
In this situation, it is obvious that 
B\"uttiker-Landauer formula could be applied. 
Our formula is applicable to general situations. 
A general discussion concerning a connection between the sign of 
$c(x)$ and the current distribution is given in Appendix C. 
It will be shown that edge current states may make a transition 
to bulk current states. 

\ 

\noindent
(6-2) Finite current effects and breakdown of QHE

Potential thus obtained modifies one-particle properties. 
Their effects become important if the magnitude of the 
total current is not infinitesimal but is finite. 
We study the system of the uniform electric field. 
Landau levels with the uniform electric field 
are not degenerate in energy and have a finite width, 
as is given in Appendix D. 
If the width in the momentum representation exceeds the Landau level 
energy spacing, QHR disappears. 
Then, breakdown of QHE occurs. 

In the von Neumann lattice representation it is easy to find the 
width in the momentum representation. 
From Eq.(D.2), the gauge invariant width is given by,
$$
eEa.
\eqno\eq
$$
QHR disappears, when the band width exceeds the Landau level's 
energy spacing, $\hbar\omega_c$, for all Landau levels 
without spin effect or for energy splitting from Zeeman energy 
due to magnetic moment $\mu$, $\mu B$. 
The critical electric field satisfies, 
$$
eE_c a=\hbar\omega_c,
\eqno\eq
$$
or 
$$
eE'_c a=\mu B, 
\eqno\eq
$$
$$
\omega_c={eB\over m^*},\ 
a=\sqrt{2\pi\hbar\over eB}=\sqrt{2\pi}l_B.
\eqno\eq
$$
where $l_B$ is the magnetic length used usually. 
Thus, the critical electric field, $E_c$ for Landau level 
splitting and $E'_c$ for spin splitting are given by, 
$$
\eqalign{
&E_c={\hbar\omega_c\over ea}=N_1 B^{3/2},\cr
&E'_c={\mu B\over ea}=N_2 B^{3/2},\cr
&N_1=25.4\times 10^3[{\rm V m^{-1} T^{-3/2}}],\cr
&N_2=0.84g\times 10^3[{\rm V m^{-1} T^{-3/2}}],
}
\eqno\eq
$$
In both cases the critical electric fields are proportional to 
$B^{3/2}$ and the numerical constants $N_1$ and $N_2$ are constant 
and independent of Landau levels. 
$g$-factor extracted from the experiment$^{\pqr}$  
is 7.3 and $N_2$ becomes 
comparable to $N_1$. 
The critical electric fields for even plateaus and odd plateaus 
are computed from the previous values as, 
$$
\eqalign{
&E_c^{\rm odd}=N^{\rm odd}B^{3/2},\cr
&E_c^{\rm even}=N^{\rm even}B^{3/2},\cr
&N^{\rm odd}=N_2=6.48\times 10^3[{\rm V m^{-1} T^{-3/2}}],\cr
&N^{\rm even}=N_1-N_2=19.0\times 10^3[{\rm V m^{-1} T^{-3/2}}]. 
}
\eqno\eq
$$
These critical fields are compared with the recent experiments of 
Kawaji et al$^\g$. 
Both of them have $B^{3/2}$ behavior and are independent from Landau 
levels. 
Our values of $N^{\rm odd}$ and $N^{\rm even}$, however, are 
much larger than the experimental observations.

The critical electric field has been estimated  before from 
naive overlappings of wave functions 
by Eaves and Sheard$^{\rt}$ and from semi-classical method by 
Trugman, and Nicopoulos and Trugman$^{\rs}$. 
In the former method, the critical electric field behaves as 
$B^{3/2}$ , and its magnitude thus obtained was similar to the 
current values. 
In the latter method, the critical electric field behaves as $B$ 
instead of $B^{3/2}$. 
We estimate the band width of the extended states due to 
Hall electric field first and compute the critical electric field 
from a condition that the bands overlapp. 
Our method is thus quite natural and leads to reasonable 
qualitative agreements,  
in the $B$-dependence and the level dependence of the critical electric 
fields with the experiments.

Edge current systems make transition to bulk current systems, from 
Appendix C. 
Consequently, the systems of only the edge currents should 
show the breakdown of the QHE in two steps. 
In the first step, an edge current system becomes to a bulk 
current system, and in the second step, 
the QHE is broken.

\chapter{Summary}

In the present work, we have shown that the quantum Hall regime(QHR) is 
realized in finite two-dimensional electron systems if the magnetic field 
is strong enough and that the quantized Hall conductance has no finite 
size correction in QHR. 
They are shown by the use of magnetic von Neumann lattice representation, 
which has been used by us before and is quite useful for studying 
one-particle properties and for showing the connection of Hall conductance 
with the topological invariant and the absence of corrections in quite 
general systems with disorders and interactions. 
The momentum representation is used, and QHR is defined based on the 
momentum representation of the propagator. 

In magnetic von Neumann lattice representation, base functions 
and dual base functions are 
local functions and have values around rectangular lattice coordinates. 
It is easy to apply the momentum representation in this method. Then 
Ward-Takahashi identity is expressed with simple and transparent form 
by use of multi-pole expansion technique. 
The derivation of the exact low energy theorem about the $\sigma_{xy}$ 
is given based on them. 
From the theorem, the $\sigma_{xy}$ at the QHR is quantized exactly as 
$(e^2/h)N$ and has no correction from finite size effect, disorders and 
interactions. 
The edge states are extended along the boundary and have continuous 
energies across Fermi energy. 
Nevertheless they have small overlapp with the momentum eigenstates, 
and the QHR is realized at the outside region of the singularities 
of the propagator in the momentum representation, $S(p)$. 
Edge states carry the electromagnetic current together with the bulk 
extended states. 
Both states contribute to Hall conductance and 
Hall conductance is quantized in QHR. 
Hall conductivity, on the other hand, may depend on spatial region 
and is not quantized generally. 

We studied the current distribution in Section 6 and found that 
the bulk states as well as the one-dimensional edge states carry the 
current generally. 
In QHR, the bulk extended states have energy gap but the edge states 
have no energy gap. 
Around Fermi energy, there are only 
one-dimensional edge states. 
B\"uttiker-Landauer formula may be applied to them, then. 

The current in QHR does not cause energy dissipation. 
One-dimensional chiral modes near Fermi energy carry a current 
without energy dissipation 
and two-dimensional extended states at the bulk with finite energy gap 
also carry a current without energy dissipation. 
In our approach, we have studied combined total current and total voltage, 
and we have shown that the $\sigma_{xy}$  are quantized exactly. 

In B\"uttiker-Landauer approach, only the states near Fermi energy are 
studied and it was shown that their contributions to $\sigma_{xy}$ 
gives the quantized $\sigma_{xy}$. 
We discussed when the current flows only in the edges. 
These systems, however, are changed to bulk current systems if 
the current exceeds the critical value.

The current in the bulk produces the Hall electric field in the bulk. 
Due to the electric field, the bulk one particle states are modified to 
become extended energetically, and QHR becomes narrow and 
eventually disappears at a critical current. 
A theoretical estimation is made. 
The results are consistent with the recent 
experiment by Kawaji et al and others$^{\g}$. 
The critical electric fields are proportional to $B^{3/2}$ and 
are independent of Landau levels in consistent with the 
experiment by Kawaji et al and others$^{\g}$, but their magnitudes 
are substantially smaller.

\ack

The present work was partially supported by the special Grant-in-Aid 
for Promotion of Education and Science in Hokkaido University Provided 
by the Ministry of Education, Science and Culture, a Grant-in-Aid for 
General Scientific Reseach (07640522), and a Grant-in-Aid for 
Scientific Reseach on Priority Area (07210202), the Ministry of 
Education, Science and Culture, Japan. 
One of the authors(K.I.) thanks Professors P. Lee, H. Nielsen, 
J. Ambjorn, I. Peterson, 
S. Kawaji, and A. Kawabata for useful discussions.

\Appendix{A}

The explicit form of matrices of Eqs.(3.5) and (3.6) are following:
$$
\eqalign{
&d_x(p)=d'_x(p)=
-i{a\over 2}({\partial\over\partial p_x}-i{\partial\over\partial p_y})
\log\alpha(p_x,p_y),\cr
&d_y(p)=d'_y(p)=
-i{a\over 2}(i{\partial\over\partial p_x}+{\partial\over\partial p_y})
\log\alpha(p_x,p_y),
}
\eqno\eq
$$
$$
\alpha(p_x,p_y)=\sum e^{ip_x(m_1-m'_1)+ip_y(n_1-n'_1)}
\langle\vec R_1\vert\vec R'_1\rangle,
\eqno\eq
$$
$$
\eqalign{
&\bar d_x(p)_{l_1,l_2}=\bar d'_x(p)_{l_1,l_2}=
{-i\over\sqrt{2eB}}(\sqrt{l_2}\delta_{l_1,l_2-1}-\sqrt{l_2+1}
\delta_{l_1,l_2+1}),\cr
&\bar d_y(p)_{l_1,l_2}=\bar d'_y(p)_{l_1,l_2}=
{1\over\sqrt{2eB}}(\sqrt{l_2}\delta_{l_1,l_2-1}+\sqrt{l_2+1}
\delta_{l_1,l_2+1}),\cr
}
\eqno\eq
$$
$\bar d_i$ satisfy the commutation relation:
$$
[\bar d_x(p),\bar d_y(p)]=[\bar d'_x(p),\bar d'_y(p)]=
-i{1\over eB}.
\eqno\eq
$$
$d_x(\vec p)$ is given in Fig.9.

\Appendix{B}

We study finite size correction of a topological invariant defined by a 
propagator, $S_{l_1,l_2}(p)$, 
$$
N_w={1\over 24\pi^2}\int d^3 p\epsilon_{\mu\nu\rho}\Tr
[{\partial S^{-1}\over \partial p_\mu}S
{\partial S^{-1}\over \partial p_\nu}S
{\partial S^{-1}\over \partial p_\rho}S].
\eqno\eq
$$
$N_w$ agrees with the integer if the momentum space is compact and the 
matrix space of $S_{l_1,l_2}(p)$ includes SU(2) group as a subspace. 
We see in the main text that the finite size correction does not 
appear in QHE. 
Finite temperature effect is similar to the finite size effect. 
$p_0$ integration in Eq.(B-1) is replaced with a discrete summation at 
finite temperature, and $N_w$ of QHE is given by, 
$$
\eqalign{
&N_w=\cases{N,&(finite size),\cr 
\sum^N_{l=1,(E_l<\mu)}(\coth{\beta\over 2}\vert\mu-E_l\vert+1)/2
,&(finite temperature),\cr}
\cr
&\beta={1\over kT},\quad\mu:{\rm chemical\ potential}.
}
\eqno\eq
$$

Finite size effect and finite temperature effect of $N_w$ depend on 
the form of the propagator. 
We compute these effects of ground state$^{\o}$ 
of Dirac field, which has 
$$
\eqalign{
&S^{-1}(p)=\gamma_\mu p^\mu+m,\cr
&\gamma_0=i\tau_3,\cr
&\gamma_1=i\tau_1,\cr
&\gamma_2=i\tau_2.
}
\eqno\eq
$$
$N_w$ is given by,
$$
N_w={1\over 2}{\vert m\vert\over m}
\cases{\coth{\vert m\vert L\over 2}
,&(finite size),\cr 
\coth{\vert m\vert\beta\over 2},&(finite temperature),\cr}
\eqno\eq
$$
where $L$ is the width in $x$-direction. 
Thus the topological invariant $N_w$ of the Dirac theory has 
the both of finite size 
correction and finite temperature correction of exponential type
$^{\p}$. 

\Appendix{C}

We study a system described by one-dimensional Landau-Ginzburg type 
static energy for $A_0$, 
$$
U=\int\nolimits^L_0 dx\{{1\over2}c(x)(\partial_x A_0)^2+{1\over4}
\lambda(x)(\partial_x A_0)^4\},
\eqno\eq
$$
with a constraint, 
$$
\sigma_{xy}I_y=\int\nolimits^L_0 dx\partial_x A_0={\rm constant}.
\eqno\eq
$$
In writting (C.1), We have ignored higher derivative terms 
such as $(\partial_x^2 A_0)^2$. 
Hence a self-consistency of Ref.18 is not considered here. 
Owing to translational invariance in $y$-direction, we use 
one-dimensional form. 
We regard the coefficient $c(x)$ and $\lambda(x)$ are known and 
study solutions of Euler-Lagrange equation under the above constraint. 
This gives the electric field and the current density. 

Depending on sign of the function $c(x)$, solutions have totally 
different properties. 

\ 

1. $c(x)>0,\ \lambda(x)>0$

Euler-Lagrange equation from (C-1) is given by,
$$
\partial_x[(\partial_x A_0)\{c(x)+\lambda(x)(\partial_x A_0)^2\}]=0. 
\eqno\eq
$$
Integrating (C-3), we have 
$$
\partial_x A_0\{c(x)+\lambda(x)(\partial_x A_0)^2\}=C_0={\rm constant}. 
\eqno\eq
$$
The electric field $\partial_x A_0$ is given by solving (C-4). 
For small $\partial_x A_0$, the solution is approximately given by, 
$$
\partial_x A_0={C_0\over C(x)}.
\eqno\eq
$$
This solution has electric field in whole region and the current flows 
in the bulk and at the edges. 
The non-zero constant $C_0$ is determined from the constraint (C-2). 

\ 

2. $c(x)>0,\ \lambda(x)>0$ in the bulk and $c(x)<0,\ \lambda(x)>0$ 
at the edges. 

Euler-Lagrange equation is the same as before and is given in (C-3). 
Because $c(x)$ is negative near the edge regions, a new type 
of solution, which corresponds to $C_0=0$, exists. 
If $C_0=0$, we have 
$$
\eqalign{
c(x)+\lambda(x)(\partial_x A_0)^2&=0, \cr
{\rm or\ }\partial_x A_0&=0.}
\eqno\eq
$$
In a region where $c(x)>0$, the electric field vanishes. 
In another region where $c(x)<0$, either the electric field vanishes 
or the electric field satisfies the first equation of (C-6). 
They are determined from the constraint of total current (C-2). 
This type of solutions have electric field only at the edge regions. 
The bulk has neither electric field nor electric current. 
Hence this corresponds to edge states. 

The edge current region where the first solution of (C-6) is 
satisfied is determined from the constraint (C-2), hence it depends 
on the total current. 
A relation between the width of the edge current region and the total 
current is given in Fig.9. 
This type of solution disappears if the current exceeds a critical 
value which satisfies, 
$$
\int\nolimits^L_{0,c(x)<0} 
dx\sqrt{-c(x)\over\lambda(x)}=\sigma_{xy}J_c.
\eqno\eq
$$
In obtaining Fig.9, we have used,
$$
\eqalign{
&\lambda(x)=\lambda_0={\rm constant}, \cr
&c(x)=\cases{c,&at the bulk,\cr 
-c+d x^2,&near the edge, $x\leq\sqrt{2c\over d}$.\cr}}
\eqno\eq
$$
The solution becomes that of $C_0\neq 0$ if the current exceeds the 
critical value, and is obtained by solving (C-4). 
It has current in the whole region. 
The current distributions of the solutions are given in Fig.10.

\Appendix{D}

In the von Neumann lattice representation, it is easy to compute 
energy spectrum in the momentum representation. 
Hamiltonian of a system with a uniform electric field, E, in the 
$x$-direction 
is given by, 
$$
\eqalign{
&H=H_0+H_1,\cr
&H_0=\sum_{l,\vec R_1}E_l b_l(\vec R_1)a_l(\vec R_1),\cr
&H_1=eE\sum b_{l_1}(\vec R_1)[{\bar d}_{x l_1,l_2}\delta_{\vec R_1,
\vec R_2}+\delta_{l_1,l_2}am_1\delta_{\vec R_1,
\vec R_2}+
\delta_{l_1,l_2}d_x(\vec R_1-\vec R_2)]
a_{l_2}(\vec R_2),}
\eqno\eq
$$
where the zero-momentum state is not included in the sets 
$\{ b_l(\vec R)\}$ and $\{ a_l(\vec R)\}$ from the constraint of the 
minimum coherent states. 
$\bar d_{xl_1 l_2}$ is given in Eq.(A.3) and the Fourier 
transform of $d_l(\vec R)$ is given in Eqs.(3.6) and (A.1). 
The first term of $H_1$ gives an inter Landau level mixings and 
the second term gives the static energy due to the electric field. 
These two terms do not contribute to the 
width of the Landau levels in the momentum representation. 
The last term, on the other hand, gives the intrinsic 
momentum dependent 
energy, $d_x(\vec p)$. 
In the momentum representation, $d_x(\vec p)$ and $d_y(\vec p)$ 
are expressed as,  
$$
\eqalign{
d_x(\vec p)=eE{a^2\over2\pi}p_y+{\partial\over\partial p_x}
\theta(\vec p), \cr
d_y(\vec p)={\partial\over\partial p_y}\theta(\vec p),\cr}
\eqno\eq
$$
with a suitable gauge function $\theta(\vec p)$ in the momentum 
space. 
A gauge invariant quantity, $\oint d_i(\vec p)dp^i$, is used for 
defining the gauge invariant energy width. 
The invariant width is given by, 
$eE{a^2\over2\pi}({2\pi\over a})=eEa$. 
Hence the Landau levels in systems of the constant electric field 
in the $x$-direction have the width, $eEa$. 
Hence the width of the extended Landau levels is inversly 
proportional to $\sqrt{B}$. 
QHR disappears if the above width agrees to the Landau level's energy 
spacing, which is proportional to $B$. 
The critical electric field of the breakdown of QHE is proportional 
to $B^{3/2}$ and the proportional constant is independent of Landau 
levels, 
in agrement with the experiments.

\FIG\fia{Energy eigenvalues (a), 
eigenfunctions (c) at $p_y=\pi/2a$, and their currents (b) are 
shown for a narrow potential well of a width, $a$. 
Left moving modes and 
right moving modes are not separated. 
In (c), potential barrier is denoted by shadow region. }

\FIG\fib{Energy eigenvalues (a), eigenfunctions (c) at 
$p_y=\pi/2a$, and their currents (b) are 
shown for a wide potential well of a width, $5a$. 
Left moving modes and 
right moving modes are separated. 
Edge states are confined in narrow edge regions. 
In (c), potential barrier is denoted by shadow region. }

\FIG\fic{Lowest order self-energy diagram which removes the 
degeneracy of Landau levels due to potentials is shown. }

\FIG\ficd{Lowest order self-energy diagram which removes the 
degeneracy of Landau levels due to interactions is shown. }

\FIG\fid{One-particle spectrum in the presence of 
the edge region is shown. 
From Eqs.(2.9), (2.11), and (2.14), in narrow regions around the 
center of Landau levels, there are extended states. 
At outside of this regions there are localized states and 
one-dimensional edge states. 
These are regarded as QHR. }

\FIG\fie{$E_F$ dependence of the topological invariant, Eq.(3.7), in 
free system is shown. }

\FIG\fif{An example of higher order diagrams is shown. 
If the internal line does not satisfy the boundary condition, 
these diagrams do not contribute to $\pi_{\mu\nu}^{(1)}(p)$ 
but contribute to $\pi_{\mu\nu}^{(2)}(p_1,p_2)$. }

\FIG\fig{Hall conductance $\sigma_{xy}$ in realistic system is given. 
In quantum Hall regimes where there are only localized states 
and one-dimensionaly extended states, 
$\sigma_{xy}$ agrees with $(e^2/h)N$. }

\FIG\fih{The width of edge current solution is obtained from the 
constraint Eq.(C-2), for the second case where $c(x)$ becomes negative 
toward the edges. 
Edge current solutions disappear if the current exceeds the 
critical value which satisfies (C.7). }

\FIG\fii{A function $c(x)$ of (C-8), and the current distribution 
for a small current case and for a large current case are shown. 
The current is restricted in narrow regions near the edges if 
$J_y<J_c$ but the current spreads into the whole region if 
$J_y>J_c$. }

\endpage
\refout
\endpage
\figout
\end